\documentclass[12pt]{article}
\usepackage[utf8]{inputenc}
\usepackage[english]{babel}
\usepackage{amsmath}
\usepackage{amsfonts}
\usepackage{amssymb}

 \usepackage{bm}

\renewcommand{\theequation}{\thesection.\arabic{equation}}

\def\lb{\label}
\def\be{\begin{equation}}
\def\ee{\end{equation}}

\begin{document}

\begin{titlepage}

\vspace*{0.7cm}

\begin{center}

{\LARGE\bf Generalization }

\vspace{0.4cm}

{\LARGE\bf of the Bargmann-Wigner construction}

\vspace{0.3cm}

{\LARGE\bf for infinite spin fields}

\vspace{1.7cm}

{\large\bf I.L.\,Buchbinder$^{1,2,3}$\!\!\!, \,\,
A.P.\,Isaev$^{2,4}$\!\!\!, \, \, M.A.\,Podoinitsyn$^{2}$\!\!, \,\,
S.A.\,Fedoruk$^2$}

\vspace{1.7cm}

\ $^1${\it Center of Theoretical Physics, Tomsk State Pedagogical University,
634061, Tomsk, Russia}, \\
{\tt joseph@tspu.edu.ru}

\vskip 0.5cm

\ $^2${\it Bogoliubov Laboratory of Theoretical Physics, JINR,
141980 Dubna, Moscow region, Russia}, \\
{\tt fedoruk@theor.jinr.ru, isaevap@theor.jinr.ru, mpod@theor.jinr.ru}

\vskip 0.5cm

\ $^3${\it Tomsk State University of Control Systems and
Radioelectronics, 634034, Tomsk, Russia}\\

\vskip 0.5cm

\ $^4${\it Lomonosov Moscow State University,
Physics Faculty, Russia} \\

\end{center}

\vspace{1.5cm}

\begin{abstract}
We develop a generalization of the Wigner scheme  for constructing
the relativistic fields corresponding to irreducible representations
of the four-dimensional Poincar\'{e} group with infinite spin. The
fields are parameterized by a vector and an additional commuting
vector or spinor variable. The equations of motion for fields of
infinite spin are derived in both formulations under consideration.
\end{abstract}

\vspace{3.5cm}

\noindent PACS: 11.10.-z, 11.30.-j, 11.30.Cp, 03.65.Pm

\vspace{0.5cm}

\noindent Keywords: unitary representations, massless infinite spin particles,  relativistic fields

\vspace{1cm}

\end{titlepage}

\setcounter{footnote}{0}
\setcounter{equation}{0}

\newpage

\setcounter{equation}0
\section{Introduction}

The classification of fields and particles and their interactions in
relativistic field theory is based on irreducible representations of
the Poincar\'{e} group or its generalizations, such as, for example,
the conformal group or supersymmetry groups. Therefore, the study of
various aspects related to space-time symmetries can lead to the
formulation of new models and approaches in classical and quantum
field theory. In this regard, the investigation of irreducible
representations of relativistic symmetry groups and their physical
applications is an important trend in modern theoretical and
mathematical physics.

The fundamentals of describing the irreducible representations of
the Poincar\'{e} group in the four-dimensional Minkowski space were
developed in the works of Wigner \cite{Wig39}, \cite{Wig48} and
Bargmann-Wigner \cite{BarWig48} (a modern state is considered e.g. in
the book \cite{BR}) . As well known, such representations are
divided into massive and massless, which are associated with massive
or massless relativistic particles\footnote{We do not discuss
physically unacceptable representations corresponding to negative
energy or imaginary mass.}. In its turn, the massless
representations are divided into representations with certain
helicities (spins) and representations with infinite (continuous)
spin. In the context of relativistic field theory, the massless
representations with definite helicities and massive representations
have been studied in sufficient detail\footnote{There is a huge
number of works devoted to descriptions of such representations. In
particular, such representations underlie the theory of higher spin
fields (see, for example, the pioneering works \cite{BarWig48},
\cite{Schwin}-\cite{F2}) or supersymmetric field theory (see e.g.
\cite{West}-\cite{BK} and references there).}. At the same time, the
representations of infinite spin, being formally physically and
mathematically consistent, did not attract attention in field
theory. However, recently there has been a surge of interest in
constructing the Lagrangian theory of infinite spin fields.
\cite{ShTor1} -- \cite{Riv2}. This emphasizes the relevance of
studying various new aspects of irreducible massless representations
of the Poincar\'{e} group and their applications.

This paper is devoted to a detailed description of the irreducible
massless representations of the four-dimensional Poincar\'{e} group with
infinite spin, the construction of infinite spin relativistic
fields, and the derivation of the equations of motion for such
fields. Relativistic fields corresponding to massless irreducible
representations with definite helicities and massive representations
of the Poincar\'{e} group were introduced in the pioneering works
\cite{Wig39}, \cite{Wig48}, \cite{BarWig48}. The equations that such
fields satisfy have different names in different contexts
(Bargmann-Wigner, Dirac-Pauli-Fierz, Rarita-Schwinger equations,
etc.) and have been studied by many authors (see, for example,
\cite{BR} and references there). In this work, we develop a
procedure for deriving the analogous equations for relativistic
fields realizing the irreducible representations of infinite spin.

The work is organized as follows. In section \ref{sect-2}, we give a
description of infinite-dimensional unitary irreducible
representations of the group $ISO(2)$ which is a small group of the
massless relativistic 4-momentum. Also, two formulations of
infinite-dimensional $ISO(2)$-representations are considered here:
one in a basis, where the states are numbered by the continuous
parameter $\varphi\in[0,2\pi)$, and second in a discrete basis. In
section \ref{sect-3}, we develop the construction of the Wigner wave
function $\Phi(p,\varphi)$ realizing a unitary irreducible the $4D$
representation of the Poincar\'{e} group of infinite spin based on the
induced transformations of the small group $ISO(2)$. We also
describe in general terms the transition scheme using the
generalized Wigner operator from the Wigner wave function
$\Phi(p,\varphi)$ to the relativistic field $\Psi(p,y)$ defined on
the space parametrized by the 4-momentum $p$ and an additional $y$
coordinate \footnote{The $y$ coordinate can be thought of as an
additional infinite-dimensional field index.}. We study two cases,
where $y$ is either a commuting 4-vector $\eta^\mu \in
\mathbb{R}^{1,3}$ or a commuting Weyl spinor $u^\alpha$, $\bar
{u}^{\do \alpha}=(u^\alpha)^*$. In the following sections, we obtain
the explicit expressions for the generalized Wigner operator in
these two cases. In section \ref{sect-4}, we consider in detail the
case when the variable $y$ is a 4-vector $\eta^\mu$. The
transformations laws of the Wigner wave function and the
relativistic field lead to some differential equations for the
Wigner operator. Solution to these equations allows to find the
Wigner operator in the explicit form. Then, using the Wigner
operator, we construct the relativistic field $\Psi(p,\eta)$ with an
additional vector variable realizing the unitary irreducible
representation of the infinite spin. The equations of motion for the
relativistic field under construction are derived. It is worth
pointing out that these equations can be treated as special case of
the Bargmann-Wigner equations \cite{Wig39,Wig48,BarWig48}. In
section \ref{sect-5}, the similar results are obtained for the case
when the variable $y$ is the Weyl spinor $u$. In addition, a
description of the relation between this formulation and the twistor
formulation of fields describing particles of infinite spin is given
here. In section \ref{sect-6} we summarize the results obtained.
Appendices 1 and 2 collect the technical details of the calculations
used in the section \ref{sect-4}.

\setcounter{equation}{0}
\section{Unitary irreducible representations of the group $ISO(2)$} \lb{sect-2}

Beginning with Wigner's paper \cite{Wig39} and subsequent papers
\cite{Wig48}, \cite{BarWig48}, it is well known that unitary
irreducible representations of the $4D$ Poincar\'{e} group are
induced by irreducible unitary representations of a small group
preserving a fixed 4-momentum of a relativistic particle in the
given representation. In the case of a massless particle in four
dimensions, a small group is the group $ISO(2)$ that is the group of
motions of the two-dimensional Euclidean space. For this reason, in
this section, we briefly discuss (for details see e.g. \cite{Vilen},
\cite{ZhSh}) the infinite-dimensional unitary representations of the
group $ISO(2)$. We describe this in a form that will be used later
in the constructing the infinite spin representations of the
Poincar\'{e} group in the four-dimensional Minkowski space.

The algebra $\mathfrak{iso}(2)$ of the group $ISO(2)$ is represented
by the Hermitian generators $T_1, T_2$ and $R$ satisfying the
following commutation relations:
\be
\lb{sope1}
[T_1,T_2] = 0 \,
,\;\;\;\;\; [R,T_a]= i\, \varepsilon_{ab} T_b \,,
\ee
where $a,b=1,2$ and $\varepsilon_{ab}$ is the antisymmetric tensor with
the components $\varepsilon_{12} =-\varepsilon_{21}=1$. An element
$h(\theta, \vec{b})$ of the $ISO(2)$ group can be written as a
product
\be
\lb{sope2}
 h(\theta, \vec{b}) = T(\vec{b}) \cdot
R(\theta) \equiv e^{  - i b_a T_a} \, e^{-i \theta R} \, ,
\ee
where $T(\vec{b})$ is the element of the translation subgroup of $ISO(2)$
and $R(\theta)$ is the element of the subgroup $SO(2)\subset
ISO(2)$. Here $\vec{b}=(b_1,b_2) \in \mathbb{R}^2$, and
$\theta \in [0,2 \pi)$ is the angular variable. The
following relation takes place
\be
\lb{sope3}
R^{-1}(\theta) \, T_a \, R(\theta)
\equiv e^{i \theta R} \, T_a \, e^{ -i \theta R} = R_{ab}(\theta)
T_b   \, ,
\ee
where $R_{ab}(\theta)$ is the rotation matrix by the
angle $\theta$ on a two-dimensional plane:
$$
||R_{ab}(\theta)|| \ = \
\begin{pmatrix}
\cos \theta & -\sin \theta \\ \sin \theta & \cos \theta
\end{pmatrix}
\,.
$$

The space $V$ of a unitary irreducible representation of the group $ISO(2)$ is spanned by the basis vectors
\be \lb{def-sp}
| \, \vec{t} \, \rangle\,, \qquad
\vec{t}=(t_1, t_2) \in \mathbb{R}^2 \,,
\ee
which are the eigenvectors for two commuting operators
$T_a$ $(a=1,2)$:
\be  \lb{sope6}
T_a | \, \vec{t} \, \rangle =
t_a | \, \vec{t} \, \rangle \, .
\ee
The action of the elements $R(\theta)\in SO(2)$ on the vectors $|\, \vec{t} \, \rangle $ is defined as
\be \lb{sope8}
T_a \, R(\theta) | \, \vec{t} \, \rangle = R(\theta) R^{-1}(\theta) \, T_a \, R(\theta) | \, \vec{t} \, \rangle  =
t'_a \, R(\theta) | \, \vec{t} \, \rangle \, .
\ee
That is, by choosing a certain normalization of the vectors $| \, \vec{t} \, \rangle$, one can put
\be \lb{sope9a}
R(\theta) \, | \, \vec{t} \, \rangle =
| \, \vec{t'} \, \rangle \; , \;\;\;\;\;\;\;\;
t_a' := R_{ab}(\theta) t_b\, ,
\ee
where the 2-vector $\vec{t'}$ is obtained by rotating $\vec{t}$ by the angle $\theta$ counterclockwise
(summation over repeated indices is implied).

The algebra $\mathfrak{iso}(2)$ has one Casimir operator
\be \lb{sope4}
T^2 := T_a T_a \, .
\ee
In the space $V$ of an irreducible representation of the group $ISO(2)$ the operator $T^2$ acts as follows
\be \lb{sope5}
T^2 = {\bm{\rho}}^2 {\cdot} I\,,
\ee
where $I$ is the identity operator, which we will omit below, and ${\bm{\rho}}$ is a fixed real number
(without loss of generality, it can be considered non-negative: ${\bm{\rho}} \geq0$). Then, due to
the relations (\ref{sope4}) and (\ref{sope5}), the irreducible unitary representation of the group $ISO(2)$ is
realized on the space $V$ with basis vectors $| \, \vec{t} \, \rangle$ from (\ref{sope6}) which lie on a
circle of radius ${\bm{\rho}}$:
\be \lb{circle}
\vec{t}{\cdot}\vec{t}:=t_a t_a = {\bm{\rho}}^2 \,.
\ee
Let us introduce the polar coordinates for such vectors\footnote{To simplify the formulas, we do not write the
radial coordinate ${\bm{\rho}}$ in the notation for the 2-vector $\vec{t}_{\varphi}$.}:
\be \lb{pol-par}
\vec{t}_\varphi := \big( (t_{\varphi})_1,(t_{\varphi})_2\big) = \left({\bm{\rho}} \cos \varphi, {\bm{\rho}} \sin \varphi \right) \,,
\ee
where $\varphi$ is an angular variable, $0\leq \varphi < 2\pi$.
Note that the transformed vector $\vec{t'}$ given in (\ref{sope9a}) has the form
\be \lb{sope9}
\vec{t'}=\vec{t}_{\varphi + \theta}\,,\qquad
\big(t_1', t_2'\big) = \big({\bm{\rho}} \cos(\varphi + \theta), {\bm{\rho}} \sin(\varphi + \theta)\big) \, .
\ee

By virtue of (\ref{pol-par}), the vectors of the irreducible representation $| \, \vec{t}_\varphi \, \rangle$
can naturally be denoted as
\be \lb{def-sp-a}
|{\bm{\rho}}, \varphi \, \rangle\,, \qquad
 |{\bm{\rho}}, \varphi +2 \pi \rangle =
 |{\bm{\rho}}, \varphi \rangle\, .
\ee
Further, for brevity, we omit the parameter ${\bm{\rho}}$ in the notation of these vectors and write $| \, \varphi \, \rangle$ instead of $| \, {\bm{\rho}}, \varphi \, \rangle$: $| \, \varphi \, \rangle \equiv |{\bm{\rho}}, \, \varphi \, \rangle$.
The completeness condition for basis vectors (\ref{def-sp-a}) is represented as
\be \lb{sop5}
\int\limits_{0}^{2 \pi} d \varphi \, |\, \varphi \, \rangle \langle \, \varphi \, | \ = \ I \,,
\ee
which corresponds to their normalization
\be \lb{norm-a}
\langle \, \varphi^{\,\prime} \, | \, \varphi \, \rangle   \ = \ \delta(\varphi-\varphi^{\,\prime}) \,,
\ee
where $\displaystyle \delta(\varphi)=\frac{1}{2\pi} \sum\limits_{n=-\infty}^{\infty}e^{\,in\varphi}$ is the
periodic $\delta $-function on a circle that satisfies the properties:
\be
\nonumber
\delta(\varphi)=\delta(\varphi+2\pi),\quad
\delta(\varphi)=\delta(-\varphi)
\ee
and for any periodic function $f(\varphi)$ we have
\be
\nonumber
\displaystyle \int\limits_{0}^{2 \pi} d \varphi^{\,\prime} \,\delta(\varphi^{\,\prime}-\varphi)
f(\varphi^{\,\prime})=f(\varphi).
\ee

Using decomposition (\ref{sope2}) for the element $h(\theta, \vec{b})$ of the group $ISO(2)$ and the
relations (\ref{sope6}), (\ref{sope8}), we find the action of the group element $ISO(2)$ on the basis vectors
\be \lb{sope10}
h(\theta, \vec{b}) | \, \varphi \, \rangle \ = \
e^{  - i \, \vec{b}_{\mathstrut} \cdot  \vec{t}_{\varphi + \theta}} | \, \varphi + \theta \, \rangle \, ,
\ee
where
\be \lb{sope10.1}
\vec{b}_{\mathstrut} \cdot  \vec{t}_{\varphi + \theta} = \rho \, \bigl( b_1 \cos (\varphi+\theta)+b_2 \sin (\varphi+\theta)\bigr) \,.
\ee
The representation (\ref{sope10}) of the group $ISO(2)$ for ${\bm{\rho}} \neq 0$ is infinite-dimensional
and using the completeness condition (\ref{sop5}), we can write
\be \lb{trans-a}
h(\theta, \vec{b}) | \, \varphi \, \rangle =
\int\limits_{0}^{2 \pi} d \varphi^{\,\prime} \,
|\, \varphi^{\,\prime} \, \rangle \langle \, \varphi^{\,\prime} \, | h(\theta, \vec{b}) | \, \varphi \, \rangle=
|\, \varphi^{\,\prime} \, \rangle \, \mathcal{D}_{\varphi^{\prime}\varphi}
(\theta, \vec{b})\, ,
\ee
where for the matrix element of the transformation we have introduced the notation
\be \lb{D-a}
\mathcal{D}_{\varphi^{\prime}\varphi}(\theta, \vec{b})
:=
\langle \, \varphi^{\,\prime} \, | h(\theta, \vec{b}) | \, \varphi \, \rangle=
e^{  - i \, \vec{b}_{\mathstrut} \cdot  \vec{t}_{\varphi^{\prime}}}\delta(\varphi^{\,\prime}-\varphi-\theta)
\ee
and "summation" over the repeated index $\varphi^{\,\prime}$ in the right-hand side (\ref{trans-a}) means
integration over $\varphi^{\,\prime}$:
\be \lb{trans-aa}
|\, \varphi^{\,\prime} \, \rangle \, \mathcal{D}_{\varphi^{\prime}\varphi}(\theta, \vec{b}) \ := \
\int\limits_{0}^{2 \pi} d \varphi^{\,\prime} \, |\, \varphi^{\,\prime} \, \rangle \,
\mathcal{D}_{\varphi^{\prime}\varphi}(\theta, \vec{b})\, ,
\ee
The action of the generators $(T_1, T_2, R)\in \mathfrak{iso}(2)$ on the basis vectors
$| \, \varphi \, \rangle$ is given by (\ref{sope6}), (\ref{sope9a}):
\be \lb{sope10.2}
T_1 | \, \varphi \, \rangle = {\bm{\rho}} \, \cos \varphi \, | \, \varphi \, \rangle  , \;\;\;
T_2 | \, \varphi \, \rangle = {\bm{\rho}} \, \sin \varphi \, | \, \varphi \, \rangle, \;\;\;
R | \, \varphi \, \rangle = i \frac{d}{d \varphi} \, | \, \varphi \, \rangle \, .
\ee
Note that the representation (\ref{trans-a}), (\ref{D-a}) is unitary, since
\be \lb{unitar}
\mathcal{D}^*_{\varphi^{\prime}\varphi}(\theta, \vec{b})
\mathcal{D}_{\varphi^{\prime}\varphi^{\prime\prime}}(\theta, \vec{b})=
\delta(\varphi-\varphi^{\prime\prime}) \, .
\ee

Using the unit decomposition (\ref{sop5}), an arbitrary vector $| \, \Phi \, \rangle$, belonging to space $V$ of
the $ISO(2)$ group unitary irreducible representation, is written in the following form:
\be \lb{sope11}
| \, \Phi \, \rangle  =
\int\limits_{0}^{2\pi} d\varphi \ \Phi(\varphi) \, | \, \varphi \, \rangle   \, ,
\ee
where the the wave function
\be \lb{sope11a}
\Phi (\varphi) := \langle \, \varphi | \, \Phi \, \rangle
\ee
is the coordinate of the vector $| \, \Phi \, \rangle$ in the given basis (\ref{def-sp-a}).
Using (\ref{sope10}), we find the action of the $ISO(2)$ group on an arbitrary element $| \, \Phi \, \rangle$
\be \lb{sope012}
| \, \Phi^\prime \, \rangle \ =  \
h(\theta, \vec{b}) \, | \, \Phi \, \rangle \  = \   \int\limits_{0}^{2\pi} d\varphi \
\Phi^\prime(\varphi) \ | \, \varphi \, \rangle\,.
\ee
Here, the transformed wave function $\Phi^\prime(\varphi)$ is given by the relation
\be \lb{sope12}
\Phi^\prime(\varphi) \ = \
\langle \, \varphi | \, h(\theta, \vec{b}) \, | \, \Phi \, \rangle  \ =  \
e^{- i \, \vec{b}_{\mathstrut} \cdot \vec{t}_\varphi} \Phi (\varphi - \theta)\,.
\ee
Contracting both parts of (\ref{sope012}) with the vector $\langle \, \varphi \, |$ on the left and using matrix elements (\ref{D-a}) as well as convention (\ref{trans-aa}) for "summation'' over the continuous index $\varphi$, we write the transformation (\ref{sope12}) as
\be \lb{sope12a}
\Phi^\prime(\varphi) \ = \ \mathcal{D}_{\varphi\varphi^{\prime}}(\theta, \vec{b}) \,\Phi (\varphi^{\prime})\,.
\ee
Then the action (\ref{sope10.2}) of $\mathfrak{iso}(2)$ generators on the wave function (\ref{sope11a}) is rewritten as follows:
\be \lb{sope13}
T_{1} \, \Phi(\varphi) = {\bm{\rho}} \, \cos\varphi \, \Phi(\varphi) \, , \quad T_{2} \, \Phi(\varphi) = {\bm{\rho}} \, \sin\varphi \, \Phi(\varphi) \, ,
\quad R \, \Phi(\varphi) = - i \frac{d}{d\varphi}  \, \Phi(\varphi)
\ee
For further it is convenient to introduce the linear combinations of generators $\mathfrak{iso}(2)$
\be \lb{sope14}
T_{\pm} := T_1 \pm i T_2 \, ,
\ee
which by virtue of (\ref{sope1}) and (\ref{sope13}) satisfy the relations
\be \lb{isou14}
[T_{+}, T_{-}] = 0\, , \qquad [R,T_{\pm}]
= \pm T_{\pm} \; ,
\ee

\be \lb{sope15}
T_{\pm} \, \Phi(\varphi) \ = \ {\bm{\rho}} \, e^{\pm i \varphi} \,  \Phi(\varphi) \, .
\ee

Sometimes it is convenient to use another basis in the space of the group $ISO(2)$ group unitary irreducible representation of the.
In this basis, the generator $R$ is diagonal while the generators of $T_a$ are not diagonal. Such a basis is
formed by a discrete set of vectors
\be \lb{basis-n}
| \, n \, \rangle\,,\qquad n \in \mathbb{Z} \, ,
\ee
and the generators of the $\mathfrak{iso}(2)$ algebra act on these vectors (\ref{basis-n}) as follows
\be \lb{sop1}
R | \, n \, \rangle = n | \, n \, \rangle \,, \qquad T_{\pm} | \, n \, \rangle = {\bm{\rho}}
| \, n \pm 1 \, \rangle \, .
\ee
Here the second relation is obtained from first one, by taking into account the commutation relation
(\ref{sope14}
 and the condition $T^2 = T_{+} T_{-}={\bm{\rho}}^ 2$, for the Casimir operator
(see (\ref{sope5})\footnote{To be more precise, the second relation in (\ref{sop1}) should
be written as
$T_{\pm} | \, n \, \rangle = e^{\pm i\alpha_n} \, {\bm{\rho}} | \, n \pm 1 \, \rangle$. However,
for what follows, we only need the fulfillment of the relation $T_{+} T_{-}={\bm{\rho}}^ 2$,
where the phase factor cancels out and we can simply omit it.}. The vectors $| \, n \, \rangle$ are orthogonal and normalized by the condition
\be \lb{norm-n}
\langle \, n \, | \, m \, \rangle = \delta_{n m} \, .
\ee
the completeness condition of the system of vectors (\ref{basis-n}), (\ref{norm-n}) has the form:
\be \lb{tcond-n}
\sum_{n=-\infty}^{\infty} | \, n \, \rangle\langle \, n \, | = 1 \, .
\ee

Transformation functions $\langle \, \varphi \, | \, n \, \rangle$ relating basis vectors (\ref{def-sp-a}) to basis vectors (\ref{basis-n}) obey the differential equation
\be \lb{sop2}
 - i \frac{d}{d\varphi} \,\langle \, \varphi \, | \, n \, \rangle = n \langle \, \varphi \, | \, n \, \rangle \, ,
\ee
which is obtained from the relations (\ref{sope10.2}) and (\ref{sop1}).
The general solution to equation (\ref{sop2}) is the $\langle \, \varphi \, | \, n \, \rangle =
c \, \exp{(i n \varphi)}$, where $c$ is some constant determined from the condition (\ref{norm-n}).
As a result, one gets
\be \lb{sop3}
\langle \, \varphi \, | \, n \, \rangle = \frac{1}{\sqrt{2\pi}}\, e^{i n \varphi}\, .
\ee

Using (\ref{tcond-n}) and (\ref{sop3}), we present the wave function $\Phi(\varphi)$ (\ref{sope11a})
as a Fourier series:
\be \lb{sope16}
\Phi(\varphi) = \langle \, \varphi \, | \, \Phi \, \rangle =\sum_{n=-\infty}^{\infty}  \langle \, \varphi \, | \, n \, \rangle \langle \, n \, | \, \Phi \, \rangle
= \sum_{n=-\infty}^{\infty} \Phi_{n} e^{i n \varphi} \, ,
\ee
where we introduced the notation
\be \lb{sop4}
\Phi_n =  \langle \, n \, | \, \Phi \, \rangle/\sqrt{2\pi}\,.
\ee
Using the polar coordinates $(b_1, b_2) = b (\cos\beta, \sin\beta)$ for the 2-vector $\vec{b}$ and relation (\ref{sope12}), we find the matrix
element of the operator $h (\theta, \vec{b})\in ISO(2)$ in the unitary representation in the discrete basis:
\be \lb{sope18}
\begin{array}{c}
\mathcal{D}_{mn} (\theta, \vec{b}) = \langle \, m \, | h(\theta, \vec{b}) | \, n \rangle =
\displaystyle
e^{- i m \theta} \frac{1}{2\pi} \int\limits_{0}^{2\pi} d\varphi \, e^{-i {\bm{\rho}} \, b \, \cos(\varphi+\theta-\beta) + i (n-m) \varphi} = \\[0.5cm]
= \displaystyle
e^{- i (n \theta - (n-m) \beta)}  \frac{1}{2\pi} \int\limits_{0}^{2\pi} d\varphi \, e^{-i {\bm{\rho}} \, b \, \cos\varphi + i (n-m) \varphi}
= \displaystyle
(- i e^{i \beta})^{n-m} \, e^{- i n \theta} J_{(n-m)}(b{\bm{\rho}}) \; ,
\end{array}
\ee
where $J_{(n)}(x)$ are the Bessel functions of integer order (Bessel coefficients). The unitarity condition for $\mathcal{D}_{mn}$ follows from the orthogonality of the Bessel coefficients
 $\sum\limits_m J_{(n-m)}^*(x) J_{(k-m)}(x)=\delta_{n,k}$.

\setcounter{equation}{0}
\section{Unitary irreducible representations of the Lorentz group} \lb{sect-3}

In this section, we briefly discuss some issues in the theory of irreducible representations of the
Lorentz and Poincar\'{e} groups in the four-dimensional Minkowski space in the form in which they will be
used below to describe relativistic fields of infinite spin.

Let us begin with reminding the notations and conventions. We will use the following Greek letters
$\mu, \nu, \lambda, \rho$ to denote the space-time indices, while the Weyl spin indices are denoted by
$\alpha$, $\beta $, $ \gamma$.

\subsection{Stability subgroup of the massless test momentum and Wigner operators}

Consider the subgroup $SL(2,\mathbb{C})\subset ISL(2,\mathbb{C})$ which acts transitively on the set ${\cal H}$ of Hermitian $(2{\times}2)$-matrices as follows (see the description of the groups $SL(2,\mathbb{C})$ and $ISL(2,\mathbb{C})$, for example, in \cite{BK,Book2,IsPod0}):
\be \lb{fr1}
A \, X \, A^{\dagger} = X' \; , \;\;\;\;\;
 \forall X \in {\cal H} \; ,
\ee
i.e. the matrix $A \in SL(2,\mathbb{C})$ maps a Hermitian matrix $X=X^\dagger$ to any other Hermitian matrix $X'=(X')^\dagger$. Since $(2{\times}2)$ matrices\footnote{We use the following set of matrices: $\sigma^0 =I_2$ and $\sigma^i$, $i=1,2,3$ -- Pauli $ \sigma$-matrices.} $\sigma^\mu=(\sigma^\mu)^\dagger$, $\mu=0,1,2,3$ form a basis in the space ${\cal H}$ , then we have the equality
\be\lb{Alambd}
A\, \sigma^\mu A^{\dagger} =
\sigma^\nu \; \Lambda_\nu^{\;\; \mu}(A)\,,
\ee
defining the matrix $||\Lambda_\nu{}^{\mu}(A)|| \in SO^{\uparrow}(1,3)$ over the element $A\in SL(2,\mathbb{C})$, and
establishing a group homomorphism $SL(2,\mathbb{C}) \to SO^{\uparrow}(1,3)$.
The expansion of the Hermitian matrix $X$ in this basis
\be\lb{herM}
X  = x_\nu \, \sigma^\nu \; ,
\ee
defines a one-to-one correspondence between the elements of ${\cal H}$ and the vectors of the four-dimensional Minkowski space\footnote{We choose the metric for $\mathbb{R}^{1,3}$ in the form $ {\mathrm{diag}}(+1,-1,-1,-1)$.} $(x_0,x_1,x_2,x_3) \in \mathbb{R}^{1,3}$.
Relations (\ref{fr1}) and (\ref{Alambd}) give
\be \lb{fr2a}
x_{\nu}' =  \Lambda_\nu^{\;\; \mu}(A) \; x_\mu  \;.
\ee
The group action\footnote{The group $ISL(2,\mathbb{C})$ covers the Poincar\'{e} group $ISO(1,3)$.} $ISL(2,\mathbb{C})$ in the space ${\cal H }$ is defined as follows: $X^\prime=AXA^\dagger +Y$, where $(A,Y)$ is an element of $ISL(2,\mathbb{C})$, where $A\in SL (2,\mathbb{C})$ and $Y\in {\cal H}$.

Let us choose the vector $\overset{_{\mathrm{\;o}}}{p} \, \in \, \overset{_{\mathstrut}}{\mathbb{R}^{1, 3}}$ as follows
\be \lb{fr3}
||\overset{_{\mathrm{\;o}}}{p}_{\nu}|| = (\overset{_{\mathrm{\;o}}}{p}_{0}, \overset{_{\mathrm{\;o}}}{p}_{1},
\overset{_{\mathrm{\;o}}}{p}_{2},
\overset{_{\mathrm{\;o}}}{p}_{3})= (E,0,0,E)
\ee
By definition, the finite-dimensional Wigner operators are the matrices $A_{(p)} \in SL(2,\mathbb{C})$ that transform the test momentum $\overset{_{\mathrm{\;o}}}{p}$ into an arbitrary momentum $p$.
That is, according to (\ref{fr1}) we have the equality
\be \lb{fr4}
A_{(p)} (\overset{_{\mathrm{\;o}}}{p}\,  \sigma) A_{(p)}^{\dagger} = (p\,\sigma) \, ,
\ee
where the notation $(p\,\sigma):= p_\mu \sigma^\mu$ is used.
The stability subgroup $G_{\overset{_{\mathrm{\;o}}}{p}}$ of the massless test momentum is formed by matrices $h \in SL(2,\mathbb{C})$ that preserve $\overset{_{\mathrm{\;o}}}{p}$:
\be \lb{fr5}
h (\overset{_{\mathrm{\;o}}}{p}\, \sigma) h^{\dagger} = (\overset{_{\mathrm{\;o}}}{p}\, \sigma) \,,
\ee
As follows from (\ref{fr4}) and (\ref{fr5}), the Wigner operators are defined up to right multiplication by elements from $G_{\overset{_{\mathrm{\;o}}}{p}}$: $A_{(p)} \simeq A_{(p)} h$. That is, the Wigner operators $A_{(p)}$ parametrize the coset space $SL(2,\mathbb{C})/G_{\overset{_{\mathrm{\;o}}}{p}}$ . To fix this parametrization, we impose the condition\footnote{We can also impose another condition for choosing elements from $SL(2,\mathbb{C})/G_{\overset{_{\mathrm{\;o}}}{p}}$, but it will be convenient for us to work with (\ref{fr6}). }
\be \lb{fr6}
A_{(\overset{_{\mathrm{\;o}}}{p})} = I \,.
\ee

The left action of an element of the group $A \in SL(2,\mathbb{C})$ on the homogeneous space $SL(2,\mathbb{C})/G_{\overset{_{\mathrm{\;o}}} {p}}$, which is parametrized by Wigner operators, is defined by the relation
\be \lb{acgf}
\Lambda \, A_{(p)} = A_{(\Lambda p)} \, h_{A,p} \, ,
\ee
where the $(4{\times}4)$-matrix of the Lorentz transformation $\Lambda$ is related to the $(2{\times}2)$-matrix $A$ via (\ref{Alambd}), and $h_ {A,p}$ is a $(2{\times}2)$-matrix from $G_{\overset{_{\mathrm{\,o}}}{p}}$. The indices of $h_{A,p}$ indicate that this matrix depends on the transformation $A \in SL(2,\mathbb{C})$ and 4-momentum $p$. Relation (\ref{acgf}) gives us expression for elements from $G_{\overset{_{\mathrm{\,o}}}{p}}$ induced by Lorentz transformations $A$:
\be \lb{acgf1}
h_{A,p} = A_{(\Lambda p)}^{-1} \, A \, A_{(p)} \;\;\;\;
\Rightarrow \;\;\;\;
h_{A,\Lambda^{-1} p} = A_{(p)}^{-1} \, A \, A_{(\Lambda^{-1} p)} \, ,
\ee
where the second expression is obtained from the first one by replacing $p$ by $\Lambda^{-1} p$.

Equation (\ref{fr5}) defining the stability subgroup $G_{\overset{_{\mathrm{\,o}}}{p}}$ of the test momentum
$\overset{_{\mathrm{\,o}} }{p}$ given in (\ref{fr3}) has the following solution
\be \lb{fr8}
h = \begin{pmatrix} e^{\frac{i}{2} \theta} & e^{-\frac{i}{2} \theta} \, \bf{b} \\ 0 & e^{-\frac{i}{2} \theta} \end{pmatrix}
=  \begin{pmatrix} 1 & \bf{b} \\ 0 & 1 \end{pmatrix} \, \begin{pmatrix} e^{\frac{i}{2} \theta} & 0  \\ 0 & e^{-\frac{i}{2} \theta} \end{pmatrix}\, ,
\ee
where $\theta \in [0,2 \pi]$ and $\mathbf{b} = b_1 + i b_2$.
This means that the matrices (\ref{fr8}) form the $ISO(2)$ group, i.e. $G_{\overset{_{\mathrm{\,o}}}{p}} \cong ISO(2)$ . The second equality in (\ref{fr8}) corresponds to the decomposition of the elements of the $ISO(2)$ group, which was considered in (\ref{sope2}). In the case of infinitesimal parameters $\theta,b_1,b_2\in \mathbb{R}$, the matrix (\ref{fr8}) has expansion
\be \lb{fr9}
h = I -i \theta \, \hat{R} -i b_1 \, \hat{T}_1 -i b_2 \, \hat{T}_2 + \dots \; ,
\ee
where the generators
\be \lb{fr10}
\hat{R} =  -\frac{1}{2}  \begin{pmatrix} 1 & 0 \\ 0 & -1 \end{pmatrix} \, , \quad
\hat{T}_1 =  \begin{pmatrix} 0 & i \\ 0 & 0 \end{pmatrix} \, , \quad
\hat{T}_2 = \begin{pmatrix} 0 & -1 \\ 0 & 0 \end{pmatrix}\,,
\ee
satisfy the relations
\be \lb{fr11}
[\hat{T}_1, \hat{T}_2] = 0 \, , \;\;\; [\hat{R}, \hat{T}_{a}] = i \varepsilon_{ad} \hat{T}_d
\ee
and are a two-dimensional realization of the real algebra $\mathfrak{iso}(2)$ with defining relations (\ref{sope1}).

The Lorentz transformations $\Lambda$ corresponding to the elements (\ref{fr8}) from the small group are determined from relation (\ref{Alambd}) for $A=h$. As a result, using equality (\ref{fr9}), we have the following three cases:
\begin{itemize}
\item  $b_1=0$, $b_2 = 0$
\be \lb{fr12}
\begin{pmatrix} e^{\frac{i}{2} \theta} & 0 \\ 0 & e^{-\frac{i}{2} \theta}  \end{pmatrix} \sigma^\mu \begin{pmatrix} e^{-\frac{i}{2} \theta} & 0 \\ 0 & e^{\frac{i}{2} \theta}  \end{pmatrix} = \sigma^\mu - i  \theta \, \sigma^\nu \mathcal{R}_\nu^{\; \mu} + \dots \, ,
\ee
\item $\theta=0$, $b_2 = 0$
\be \lb{fr13}
\begin{pmatrix} 1 & b_1 \\ 0 & 1  \end{pmatrix} \sigma^\mu \begin{pmatrix} 1 & 0 \\ b_1 & 1  \end{pmatrix} = \sigma^\mu - i  b_1  \, \sigma^\nu (\mathcal{T}_1)_\nu^{\; \mu} + \dots \, ,
\ee
\item $\theta = 0$, $b_1 = 0$
\be \lb{fr14}
\begin{pmatrix} 1 & i b_2 \\ 0 & 1  \end{pmatrix} \sigma^\mu \begin{pmatrix} 1 & 0 \\ -i b_2 & 1  \end{pmatrix} = \sigma^\mu - i  b_2  \, \sigma^\nu (\mathcal{T}_2)_\nu^{\; \mu} + \dots \, ,
\ee
\end{itemize}
where the matrix generators
\be \lb{fr15}
|| \mathcal{R}_\nu^{\; \mu} || = \begin{pmatrix} 0 & 0 & 0& 0 \\ 0 & 0 & i & 0 \\ 0 & -i & 0 & 0 \\ 0 & 0 & 0 & 0 \end{pmatrix} , \;\;
|| (\mathcal{T}_1)_\nu^{\; \mu}|| = \begin{pmatrix} 0 & i & 0 & 0 \\ i & 0 & 0 & -i \\ 0 & 0 & 0 & 0 \\ 0 & i & 0 & 0 \end{pmatrix} , \;\;
|| (\mathcal{T}_2)_\nu^{\; \mu}|| = \begin{pmatrix} 0 & 0& -i & 0 \\ 0 & 0 & 0 & 0 \\ -i & 0 & 0 & i \\ 0 & 0 & -i & 0 \end{pmatrix}
\ee
realize the four-dimensional representation of the algebra $\mathfrak{iso}(2)$ with the defining relations (\ref{sope1}):
\be \lb{fr16}
[\mathcal{T}_1, \mathcal{T}_2] = 0 \, , \;\;\; [\mathcal{R}, \mathcal{T}_a] = i \varepsilon_{ad} \mathcal{T}_d \, .
\ee
Thus, in the case of infinitesimal parameters, the element (\ref{fr8}) of the small group $h$ corresponds to the four-dimensional Lorentz transformation
\be \lb{lsh}
\Lambda_\nu^{\; \mu}(h)  = \delta_\nu^{\; \mu} -i  \theta \, \mathcal{R}_\nu^{\; \mu} -i b_1  \, (\mathcal{T}_1)_\nu^{\; \mu} -i  b_2  \, (\mathcal{T}_2)_\nu^{\; \mu} + \dots \, ,
\ee
where the matrices $\mathcal{R}$, $\mathcal{T}_1$, $\mathcal{T}_2$ are given in (\ref{fr15}).

\subsection{Wigner wave functions and massless relativistic fields}

Relativistic states of zero mass are given by vectors in the space of the $ISO(2)$-representation in a fixed reference frame of the test momentum, which in the momentum representation are functions of the energy-momentum vector $p_\mu$. Consequently, the basis in the state space of relativistic massless particles with light-like momentum $p\in \mathbb{R}^{1,3}$ (on which the unitary irreducible representation of the group $SL(2,\mathbb{C})$ is realized as induced from the unitary representation $ISO(2)$) is given by the vectors
\be \lb{r-a}
|\,p, \varphi \, \rangle\, ,\qquad \varphi \in [0,2\pi].
\ee
The basis can also be given in discrete form in terms of the vectors
\be \lb{r-n}
|\, p, n \, \rangle \, ,\qquad n\in \mathbb{Z}\,.
\ee
The vectors (\ref{r-a}),\, (\ref{r-n}) generalize the states (\ref{def-sp-a}) and (\ref{basis-n}). Since we are studying the massless case, the 4-momentum $p$ in (\ref{r-a}) and (\ref{r-n}) obeys the conditions $p^2=0$, $p_0\geq0$.

Thus, the unitary representation of the covering $SL(2, \mathbb{C})$ of the Lorentz group for the element
$A \in SL(2, \mathbb{C})$ is found by the method of induced representa\-tions, it is induced from the
representa\-tions of the stability(small) group of the standard test momentum. In this construction,
the unitary representation of the element $A \in SL(2, \mathbb{C})$ is given by transforming the vectors $|\, p, \varphi \, \rangle$:
\be \lb{trans-rs}
U(A) |  \, p, \varphi \, \rangle =
|\  \Lambda p , \varphi^{\,\prime} \, \rangle \, \mathcal{D}_{\varphi^{\prime}\varphi}(\theta_{A,p}, \vec{b}_{A,p})\, ,
\ee
or in the explicit form
\be \lb{unpgr1}
U(A) | \, p, \varphi \, \rangle = h(\theta_{A,p}, \vec{b}_{A,p}) | \Lambda p , \varphi \, \rangle \, =
e^{- i \vec{b}_{A,p} \cdot \vec{t}_{\varphi+\theta_{A,p}}} | \  \Lambda p , \varphi+\theta_{A,p} \, \rangle \, .
\ee
These relations follow directly from expressions (\ref{trans-a}), (\ref{D-a}). Here the operators
$h(\theta_{A,p}, \vec{b}_{A,p})$ are elements of the small group $ISO(2)$ in the unitary irreducible
representation (\ref{sope10}), which was described in the previous section. The matrix elements
$\mathcal{D}_{\varphi^{\prime}\varphi}(\theta_{A,p}, \vec{b}_{A,p})$ of these operators are defined by
the relation (\ref{D-a}). It is important to note that now the parameters $\theta$ and $\vec{b}$ of the group
$ISO(2)$ depend on the element $A \in SL(2,\mathbb{C})$ and the four-dimensional momentum $p$. This fact is
indicated through the notation $\theta_{A,p}, \vec{b}_{A,p}$. The parameters $\theta_{A,p}$ and $ \vec{b}_{A,p}$
correspond to the element $h_{A,p}$, which is given by the first relation it (\ref{acgf1}). As before, the matrix $\Lambda$ is related to $A$ by (\ref{Alambd}). We also give an explicit expression for the exponent in (\ref{unpgr1}):
\be \lb{exf}
\vec{b}_{A,p} \cdot \vec{t}_{\varphi+\theta_{A,p}} = {\bm{\rho}} \big[ (b_{A,p})_1 \, \cos (\varphi+\theta_{A,p}) + (b_{A,p})_2 \,
\sin (\varphi+\theta_{A,p}) \big] \,  ,
\ee
where the non-negative real parameter ${\bm{\rho}}$ characterizes the unitary irreducible representation of the
small group $ISO(2)$, and thus characterizes the unitary irreducible massless representation of the
covering $ISL(2,\mathbb{C})$ Poincar\'{e} groups.

The unitary representation of the covering Lorentz group $SL(2,\mathbb{C})$, realized on the basis vectors
$|\, p, \varphi \, \rangle$, can be easily rewritten in the discrete basis $|\, p, n \, \rangle$. In this case,
the matrix elements of the small group are determined by relation (\ref{sope18}).

An arbitrary vector in the space of the massless unitary irreducible representation of the group $SL(2,\mathbb{C})$ is a linear combination of the basis vectors (\ref{r-a})
\be \lb{unpgr2}
| \, \Phi \, \rangle \ = \ \int d^{\,4} p \ \vartheta(p_0)\delta(p^2)\int\limits_{0}^{2 \pi} d \varphi \ \Phi(p,\varphi)\, |\, p, \varphi \, \rangle  \, ,
\ee
where $\vartheta(p_0)$ is the Heaviside function: $\vartheta(p_0)=1$ for $p_0\geq0$ and $\vartheta(p_0)=0$ for $p_0<0$.
The "coordinate" functions $\Phi(p,\varphi)$ in this integral expansion will be called the Wigner wave functions of a massless particle.
These functions are obtained by inducing from the wave functions (\ref{sope11a}).
Note that instead of the decomposition (\ref{sope16}), we can use the decomposition
\be \lb{Phi-exp}
\Phi(p,\varphi)
= \sum_{n=-\infty}^{\infty} \Phi_{n}(p)
 e^{i n \varphi} \, ,
\ee
where $\Phi_{n}(p)$ are the functions of the momentum variable $p=(p_0,p_1,p_2,p_3)$.

The induced unitary representations of the group $SL(2,\mathbb{C})$ realized on the Wigner functions $\Phi(p,\varphi)$ are constructed according to (\ref{sope12a}), (\ref{trans-rs}) and (\ref{unpgr1}) and have the form:
\be \lb{trans-Wvf}
\Phi^\prime(p,\varphi) \ := \ [U(A) \Phi](p,\varphi) \ = \  \mathcal{D}_{\varphi\varphi^{\prime}}(\theta_{A,\Lambda^{-1} p}, \vec{b}_{A,\Lambda^{-1} p}) \,
\Phi (\Lambda^{-1} p , \varphi^{\prime})\,,
\ee
i.e.
\be \lb{unpgr3}
\Phi^\prime(p,\varphi) \ = \ e^{- i \vec{b}_{A,\Lambda^{-1}  p} \cdot \vec{t}_{\varphi}} \Phi( \Lambda^{-1} p ,
\varphi - \theta_{A,\Lambda^{-1} p})  \, .
\ee

Note that the transformations of the Wigner functions (\ref{trans-Wvf}), (\ref{unpgr3}) depend on the momentum variable $p_\mu$. Our goal is to find relativistic fields on which unitary irreducible representations of the Lorentz and Poincar\'{e} groups are realized, and the transformations themselves do not depend on the 4-momentum and are determined only by the matrix $A\in SL(2,\mathbb{C})$.

Let us assume that the Lorentz-covariant field $\Psi(p, y)$, which describes massless particles, is constructed from the Wigner function $\Phi(p,\varphi)$ using the integral transformation
\be \lb{unpgr4}
\Psi(p, y) \ = \ \int\limits_{0}^{2\pi} d \varphi \, \mathcal{A}(p,y,\varphi)\, \Phi (p, \varphi) \, ,
\ee
where $y$ is some set of auxiliary variables.
The function $\mathcal{A}(p,y,\varphi)$ plays the role of the kernel of the Wigner operator, which is an
infinite-dimensional analogue of the operator $A_{(p)}$ from (\ref{fr4}). If we use the convention
(\ref{trans-aa}) on summation (integration) over the repeated index $\varphi$, then expression (\ref{unpgr4})
is written in short form
\be \lb{unpgr4a}
\Psi(p, y) \ = \ \mathcal{A}(p,y,\varphi)\, \Phi (p, \varphi) \, .
\ee
The choice of auxiliary variables $y$ seems to be rather arbitrary. The main requirement for the field
$\Psi(p, y)$ is that it has standard Lorentz transformations independent of the momentum variables $p$.
This condition somewhat restricts the choice of variables $y$ and for each such choice will lead to a certain
form of the kernel $\mathcal{A}(p,y,\varphi)$. In a sense, the set of variables $y$ can be considered as an
additional index of the field $\Psi(p, y)$. Then, the transformation (\ref{unpgr4a}) is interpreted as a
transition using the Wigner operator $\mathcal{A}(p,y,\varphi)$ from the wave function $\Phi (p, \varphi)$ to
the field $\Psi( p, y)$ with the help of reduction by index $\varphi$. Such a transition for massless
representations was proposed in \cite{Weinb1}, \cite{Weinb2}. For massive representations $ISL(2,\mathbb{C})$
a similar approach and Wigner operators were considered in detail in \cite{IsPod0}. For massless
representations with definite helicities similar constructions were described in \cite{ZFed}.

When considering the massless representations of infinite (continuous) spin, the use of additional variables in
the field description is quite natural since these representations must describe an infinite number of spin
states. The commuting vector $\eta_\mu=(\eta_0,\eta_1,\eta_2,\eta_3)$
was used as auxiliary variables for various special purposes in \cite{ShTor1}, \cite{ShTor2}, and in
\cite{BFIR}, \cite{BFI} the commuting Weyl spinor $u^\alpha=(u^1,u^2)$ was used. It is clear that other
choices of auxiliary variables $y$ are also possible.

Further, we describe the construction of relativistic fields of infinite spin for both of the above cases of
additional variables and construct the equations of motion for such fields.

\setcounter{equation}{0}
\section{Relativistic fields with an additional vector variable} \lb{sect-4}

Consider the case when the commuting 4-vector $\eta=(\eta_0,\eta_1,\eta_2,\eta_3)\in\mathbb{R}^{1,3}$ is taken
as a set of variables $y$ (see \cite{Wig48}, \cite{BarWig48}).
In this section, we follow the approach developed in \cite{ShTor1}, \cite{ShTor2}.

Let the relativistic field $\Psi(p, \eta)$ given in (\ref{unpgr4}), (\ref{unpgr4a}) be transformed under the action of the Lorentz group in the standard way:
\be \lb{unpgr41}
\Psi^\prime(p, \eta) \ = \ [U(A) \Psi](p, \eta) \ = \ \Psi(\Lambda^{-1} p , \Lambda^{-1} \eta) \, ,
\ee
where the matrices $A$ and $\Lambda$ are related by (\ref{Alambd}). Knowing the explicit form of the unitary
Lorentz transformation (\ref{trans-Wvf}), (\ref{unpgr3}) of the wave function $\Phi(p, \varphi)$ and the corresponding transformation (\ref{unpgr41}) of the field $\Psi( p, \eta)$, we find the equations that determine the kernel $\mathcal{A}(p,\eta,\varphi)$ of the Wigner operator present in expressions (\ref{unpgr4}), (\ref{unpgr4a}).

First, from (\ref{unpgr41}) and (\ref{unpgr4}) one gets the equality
\be \lb{unpgr5}
[U(A) \Psi](p,\eta) = \Psi(\Lambda^{-1} p , \Lambda^{-1} \eta) =  \int\limits_{0}^{2\pi} d \varphi \, \mathcal{A}( \Lambda^{-1} p, \Lambda^{-1}\eta,\varphi) \,\Phi (\Lambda^{-1} p, \varphi)  \, .
\ee
On the other hand, we apply the transformation law (\ref{unpgr3}) to (\ref{unpgr4}) and obtain:
\begin{eqnarray}
\nonumber
[U(A) \Psi](p,\eta) &=& \int\limits_{0}^{2\pi} d \varphi \, \mathcal{A}(p, \eta,\varphi)\, [U(A) \Phi] (p, \varphi) \\ [7pt]
\nonumber
&=& \int\limits_{0}^{2\pi} d \varphi \, \mathcal{A}(p, \eta,\varphi) e^{- i \vec{b}_{A,\Lambda^{-1} p} \vec{t}_{\varphi}}  \, \Phi( \Lambda^{-1} p , \varphi - \theta_{A,\Lambda^{-1} p}) \\ [7pt]
\lb{unpgr6}
&=&  \int\limits_{0}^{2\pi} d \varphi \, \mathcal{A}(p, \eta,\varphi+ \theta_{A,\Lambda^{-1} p}) e^{- i \vec{b}_{A,\Lambda^{-1} p}  \vec{t}_{\varphi+\theta_{A,\Lambda^{-1} p}}} \, \Phi( \Lambda^{-1} p , \varphi) \, .
\end{eqnarray}
Equating the relations (\ref{unpgr5}) and (\ref{unpgr6}), we get the following equation for the kernel $\mathcal{A}(p,\eta,\varphi)$:
\be \lb{unpgr7}
\mathcal{A}(\Lambda^{-1} p, \Lambda^{-1} \eta, \varphi) \ = \ e^{- i \vec{b}_{A,\Lambda^{-1} p} \vec{t}_{\varphi+\theta_{A,\Lambda^{-1} p}}} \mathcal{A}(p,\eta,\varphi+\theta_{A,\Lambda^{-1} p})\,.
\ee
Write the right-hand side (\ref{unpgr7}) as
\be \lb{unpgr7a}
e^{- i \vec{b}_{A,\Lambda^{-1} p} \vec{t}_{\varphi+\theta_{A,\Lambda^{-1} p}}} \mathcal{A}(p,\eta,\varphi+\theta_{A,\Lambda^{-1} p})
 \ = \  \mathcal{A}(p,\eta,\varphi^\prime)\,
 \mathcal{D}_{\varphi^\prime\varphi}(\theta_{A,\Lambda^{-1} p},\vec{b}_{A,\Lambda^{-1} p})\,,
\ee
where $\mathcal{D}_{\varphi^\prime\varphi}$ is the matrix element given in (\ref{D-a}).
Rewriting (\ref{unpgr7}) and taking into account (\ref{unpgr7a}), one gets the relation
\be \lb{unpgr7aa}
\mathcal{A}(\Lambda^{-1} p, \Lambda^{-1} \eta, \varphi) \ = \
\mathcal{A}(p,\eta,\varphi^\prime)\,
\mathcal{D}_{\varphi^\prime\varphi}(\theta_{A,\Lambda^{-1} p},\vec{b}_{A,\Lambda^{-1} p}) \, ,
\ee
which is an infinite-dimensional analogue of equality (\ref{acgf}) for the Wigner operator $A_{(p)}$. We see that
the kernel $\mathcal{A}(p,\eta,\varphi)$ plays the role of the matrix $A_{( p)}$.

Equations (\ref{unpgr7}), (\ref{unpgr7a}) lead to two important consequences for the kernel
$\mathcal{A}(p,\eta,\varphi)$ of the Wigner operator.

First, let us take in (\ref{unpgr7}) as the Lorentz transformation $A$ in the form $A = A_{(p)}$.
Taking into account the relation
\be \lb{sur1}
\Lambda^{-1}(A_{(p)}) p =
 \overset{_{\mathrm{\;o}}}{p} \, ,
\ee
and the conditions (\ref{fr6}) one sees that the second relation in (\ref{acgf1}) leads to $h_{A_{(p)},p}=1$ and all parameters of the stability subgroup
element on the right-hand side of equality (\ref{unpgr7}) are zero. Thus, for $A = A_{(p)}$, the relations
(\ref{unpgr7}), (\ref{unpgr7aa}), after rearranging the left-hand side and right-hand side, take the form:
\be \lb{unpgr8}
\mathcal{A}(p,\eta,\varphi) \ = \ \mathcal{A}(\overset{_{\mathrm{\;o}}}{p}, \Lambda^{-1}(A_{(p)}) \eta, \varphi)  \, .
\ee
That is, to construct the kernel $\mathcal{A}(p,\eta,\varphi)$ for an arbitrary momentum $p$, it suffices to
find the kernel $\mathcal{A}(\overset{_{\mathrm{\;o}}}{p},\eta,\varphi)$
at the test momentum $\overset{_{\mathrm{\;o}}}{p}$, i.e $\mathcal{A}(p,\eta,\varphi) $ is obtained from $\mathcal{A}(\overset{_{\mathrm{\;o}}}{p},\eta,\varphi)$ using the Lorentz transformation of the vector $\eta$.

The second corollary from relation (\ref{unpgr7}) leads to the equation that defines the kernel $\mathcal{A}(\overset{_{\mathrm{\;o}}}{p},\eta,\varphi )$.
To do this, we first put
$p = \overset{_{\mathrm{\;o}}}{p}$, and then take $A = h$ on the right-hand side and left-hand side of relation (\ref{unpgr7}) we first put
$p = \overset{_{\mathrm{\;o}}}{p}$. Here the element $h \in ISO(2)$ is taken in the
parameterization (\ref{fr8})
and depends on $\vec{b}$ and $\theta$. As a result, one gets the following relation:
\be \lb{unpgr9}
\mathcal{A}(\overset{_{\mathrm{\;o}}}{p}, \Lambda^{-1}(h) \eta, \varphi) = e^{- i \vec{b} \cdot \vec{t}_{\varphi+\theta}} \mathcal{A}(\overset{_{\mathrm{\;o}}}{p},\eta,\varphi+\theta)
\, ,
\ee
where the matrix $\Lambda(h)$ in the left-hand side of equality (\ref{unpgr9}) is defined by (\ref{Alambd})
and is given in infinitesimal form by relation (\ref{lsh}).

Equation (\ref{unpgr9}) shows that  the representation of the three-parameter small group $ISO(2)$ is realized on the kernel $\mathcal{A}(p, \eta, \varphi)$ for $p=\overset{_{\mathrm{\;o}}}{p} $.
Considering the corresponding infinitesimal values of the parameters and using the expansion (\ref{lsh})
on the left side (\ref{unpgr9}), we obtain three $iso(2)$-algebraic relations, which are written as three
differential equations for the kernel $\mathcal {A}(\overset{_{\mathrm{\;o}}}{p},\eta,\varphi)$:
\begin{itemize}
\item When $\theta$ is small, $b_1=0$, $b_2 = 0$, the equation (\ref{unpgr9}) becomes
\be \lb {fe1a}
\mathcal{A}(\overset{_{\mathrm{\;o}}}{p}, \eta_\mu + i  \theta \, \mathcal{R}_{\mu}{}^{\nu} \eta_\nu - \dots , \varphi) =\mathcal{A}(\overset{_{\mathrm{\;o}}}{p},\eta,\varphi)+ \theta \, \frac{\partial}{\partial \varphi}  \, \mathcal{A}(\overset{_{\mathrm{\;o}}}{p},\eta,\varphi)+ \dots
\ee
Expanding the left-hand side (\ref{fe1a}) into a series on the small parameter $\theta$ and using (\ref{fr15}),
we obtain
\be \lb {fe2a}
\left( \eta_1 \frac{\partial}{\partial \eta^{2}} - \eta_2 \frac{\partial}{\partial \eta^{1}} \right)
\mathcal{A}(\overset{_{\mathrm{\;o}}}{p},\eta,\varphi) = - \frac{\partial}{\partial \varphi}  \, \mathcal{A}(\overset{_{\mathrm{\;o}}}{p},\eta,\varphi)\,.
\ee

\item In the case of $\theta = 0$, $b_1$ is small, $b_2 = 0$, the equation (\ref{unpgr9}) becomes
\be \lb {fe1b}
\mathcal{A}(\overset{_{\mathrm{\;o}}}{p}, \eta_\mu + i  b_1 \, (\mathcal{T}_1)_{\mu}{}^{\nu} \eta_\nu - \dots , \varphi) =\mathcal{A}(\overset{_{\mathrm{\;o}}}{p},\eta,\varphi) - i {\bm{\rho}} \, b_1 \, \cos \varphi \, \mathcal{A}(\overset{_{\mathrm{\;o}}}{p},\eta,\varphi)+ \dots
\ee
After using (\ref{fr15}), in the first order in the parameter $b_1$ we get
\be \lb{fe2b}
\left[ \eta_1 \left(\frac{\partial}{\partial \eta^0} - \frac{\partial}{\partial \eta^3} \right)- (\eta_0 - \eta_3) \frac{\partial}{\partial \eta^1} \right] \mathcal{A}(\overset{_{\mathrm{\;o}}}{p},\eta,\varphi) = i {\bm{\rho}} \, \cos \varphi \, \mathcal{A}(\overset{_{\mathrm{\;o}}}{p},\eta,\varphi)\,.
\ee

\item In the case of $\theta = 0$, $b_1 = 0$, $b_2$ is small, the equation (\ref{unpgr9}) implies the equation
\be \lb {fe1c}
\mathcal{A}(\overset{_{\mathrm{\;o}}}{p}, \eta_\mu + i  b_2 \, (\mathcal{T}_2)_{\mu}{}^{\nu} \eta_\nu - \dots , \varphi) =\mathcal{A}(\overset{_{\mathrm{\;o}}}{p},\eta,\varphi)-i  {\bm{\rho}} \, b_2 \, \sin \varphi \, \mathcal{A}(\overset{_{\mathrm{\;o}}}{p},\eta,\varphi)+ \dots,
\ee
which in the first order with respect to the parameter $b_2$ is written as
\be \lb{fe2c}
\left[(\eta_0 - \eta_3) \frac{\partial}{\partial \eta^2} - \eta_2 \left(\frac{\partial}{\partial \eta^0} - \frac{\partial}{\partial \eta^3} \right) \right] \mathcal{A}(\overset{_{\mathrm{\;o}}}{p},\eta,\varphi) = i {\bm{\rho}} \, \sin \varphi \, \mathcal{A}(\overset{_{\mathrm{\;o}}}{p},\eta,\varphi)
\ee
after using (\ref{fr15}).
\end{itemize}

The resulting equations (\ref{fe2a}), (\ref{fe2b}), (\ref{fe2c}) are necessary equations that the kernel
$\mathcal{A}(\overset{_{\mathrm{\; o}}}{p},\eta,\varphi)$ must satisfy. As will be shown, executing these
equations for $\mathcal{A}(\overset{_{\mathrm{\;o}}}{p},\eta,\varphi)$ will result to the required kernel
$\mathcal{A} (p,\eta,\varphi)$, which is defined by relation (\ref{unpgr8}). Then with the help of
(\ref{unpgr4}), (\ref{unpgr4a}) we can construct the field $\Psi(p, \eta)$.

Equations (\ref{fe2a}), (\ref{fe2b}), (\ref{fe2c}) have two types of solutions depending on the value
of $\eta^+:=\eta^0+\eta^3$. In the case of $\eta^+\neq0$, we obtain a solution called in \cite{ShTor1,ShTor2,ShTorZh} non-singular, while for $\eta^+=0$ there arise solutions with additional $\delta$-functions, which, according to the terminology of \cite{ShTor1,ShTor2,ShTorZh}, are called singular. In the next subsection, we describe these two classes of solutions.

\subsection{Non-singular solution}

As shown in Appendix 1, the solution of equations (\ref{fe2a}), (\ref{fe2b}), (\ref{fe2c}) for the kernel at $\eta^+\neq0$ is the expression
\be \lb{sol}
\mathcal{A}(\overset{_{\mathrm{\;o}}}{p},\eta,\varphi) = e^{i {\bm{\rho}} \,\left(\eta_1 \cos \varphi - \eta_{2} \sin \varphi\right)/{\eta^+}  } \, f(\eta \cdot \eta , \eta^+) \, ,
\ee
where $f((\eta)^2, \eta^+)$ is an arbitrary function of $(\eta)^2 := \eta \cdot \eta$ and $\eta^+=\eta_0 - \eta_3 $. Here and below we use the notation $(a \cdot b)$ for the scalar product of any two vectors $a,b \in \mathbb{R}^{1,3}$.

The solution (\ref{sol}) can also be written as
\be \lb{sol1}
\mathcal{A}(\overset{_{\mathrm{\;o}}}{p},\eta,\varphi) =
e^{i {\bm{\mu}} \, ( \eta \cdot \overset{_{\mathrm{\,o}}}{\varepsilon}_{(1)} \, \cos \varphi - \eta \cdot \overset{_{\mathrm{\,o}}}{\varepsilon}_{(2)} \, \sin \varphi) / (\eta\cdot \overset{_{\mathrm{\;o}}}{p} )}  \, f(\eta \cdot \eta , \eta \cdot \overset{_{\mathrm{\;o}}}{p}) \, ,
\ee
where we have used the relation (\ref{fr3}). Instead of the dimensionless constant ${\bm{\rho}}$, we
introduced the mass dimensional constant
\be \lb{mu}
{\bm{\mu}}:= E{\bm{\rho}} \, ,
\ee
instead of an arbitrary function $f((\eta)^2 , \; \eta \cdot \overset{_{\mathrm{\;o}}}{p}/E)$,
we introduced the function $f( (\eta)^2 , \, \eta \cdot \overset{_{\mathrm{\;o}}}{p})$, and also we introduced
two additional 4-vectors with the coordinates
\be \lb{sol2}
(\overset{_{\mathrm{\,o}}}{\varepsilon}_{(1)})_\nu = (0,1,0,0) \, , \qquad
(\overset{_{\mathrm{\,o}}}{\varepsilon}_{(2)})_\nu = (0,0,1,0) \, .
\ee
Let us define the vectors
\be \lb{sol2s}
\overset{_{\mathrm{\,o}}}{\varepsilon}_{(1)}(\varphi)
:=\overset{_{\mathrm{\,o}}}{\varepsilon}_{(1)}\cos\varphi- \overset{_{\mathrm{\,o}}}{\varepsilon}_{(2)}\sin\varphi\,,\qquad
\overset{_{\mathrm{\,o}}}{\varepsilon}_{(2)}(\varphi)
:=\overset{_{\mathrm{\,o}}}{\varepsilon}_{(1)}\sin\varphi+ \overset{_{\mathrm{\,o}}}{\varepsilon}_{(2)}\cos\varphi\,,
\ee
that are $SO(2)$-transformations of $\overset{_{\mathrm{\,o}}}{\varepsilon}_{(1)}$ and $\overset{_{\mathrm {\,o}}}{\varepsilon}_{(2)}$.
Then the expression (\ref{sol1}) takes the form
\be \lb{sol1a}
\mathcal{A}(\overset{_{\mathrm{\;o}}}{p},\eta,\varphi) = e^{i {\bm{\mu}} \, \eta\cdot \overset{_{\mathrm{\,o}}}{\varepsilon}_{(1)}(\varphi)  /
(\eta\cdot \overset{_{\mathrm{\;o}}}{p} )}  \, f(\eta \cdot \eta ,  \,
\eta\cdot \overset{_{\mathrm{\;o}}}{p}) \, .
\ee

The expression for the kernel $\mathcal{A}(p,\eta,\varphi)$ in the case of an arbitrary momentum $p$
is obtained from expression (\ref{sol}) by means of relation (\ref{unpgr8}). Using (\ref{sur1}), we get the following expression:
\be \lb{sol3}
\mathcal{A}(p,\eta,\varphi) = e^{\,i {\bm{\mu}} \,  \eta\cdot \varepsilon_{(1)}(\varphi) /(\eta \cdot p)} \, f(\eta \cdot \eta,  \eta \cdot p) \, .
\ee
Also, in (\ref{sol3}) we used the notation for one of $SO(2)$-rotated polarization vectors
\be \lb{sol4}
\varepsilon_{(1)}(\varphi) = \Lambda(A_{(p)}) \; \overset{_{\mathrm{\,o}}}{\varepsilon}_{(1)}(\varphi) \, , \qquad
\varepsilon_{(2)}(\varphi) = \Lambda(A_{(p)}) \; \overset{_{\mathrm{\,o}}}{\varepsilon}_{(2)}(\varphi) \, ,
\ee
that are orthogonal to each other and transverse to the massless four-momentum $p$:
\be \lb{sor1}
\varepsilon_{(1)}(\varphi)  \cdot p= \varepsilon_{(2)}(\varphi) \cdot p = 0  \, , \qquad
\varepsilon_{(1)}(\varphi)  \cdot \varepsilon_{(2)}(\varphi)  = 0 \,.
\ee
These polarization vectors are normalized as follows:
\be \lb{sor2}
\varepsilon_{(1)}(\varphi)  \cdot \varepsilon_{(1)}(\varphi) = \varepsilon_{(2)}(\varphi)  \cdot \varepsilon_{(2)}(\varphi)  = -1\,.
\ee
The solution (\ref{sol3}) reproduces the solution found in \cite{ShTor1}, \cite{ShTor2}.

Using the obtained expression for the generalized Wigner operator (\ref{sol3}) and the relation (\ref{unpgr4})
we can construct the Lorentz-covariant field $\Psi(p,\eta)$. As a result, one gets
\be \lb{nrez1}
\Psi(p,\eta) = \int\limits_{0}^{2\pi} d \varphi \,  e^{i {\bm{\mu}} \, \eta\cdot \varepsilon_{(1)}(\varphi)/ ( \eta \cdot p) } \,
f(\eta \cdot \eta, \eta \cdot p) \,
\Phi (p, \varphi) \, .
\ee
Let us introduce the Pauli-Lubanski vector
\be
\lb{vPL}
\hat{W}_\mu=\frac12\,\varepsilon_{\mu\nu\lambda\rho}\hat{P}^\nu \hat{M}^{\lambda\rho} \; ,
\ee
where the operators of the components of the momentum $\hat{P}^\nu$ and angular momentum
$\hat{M}^{\lambda\rho}$ are given in Appendix 2. In Appendix 2, we showed that on the field (\ref{nrez1})
the square of the Pauli-Lubanski vector takes a fixed value $-{\bm{\mu}}^2$. This means that the fields
$\Psi(p, \eta)$ constructed on the base of Wigner wave functions $\Phi (p, \varphi)$ corresponds to
irreducible representation of the covering group $ISL(2,\mathbb{C})$ of infinite spin.

Using the expansion (\ref{Phi-exp}) for the Wigner wave function $\Phi(p,\varphi)$, we rewrite the expression (\ref{nrez1}) as
\be \lb{nrez-ra}
\Psi(p,\eta) = f(\eta \cdot \eta, \eta \cdot p) \sum_{n\in \mathbb{Z}} \int\limits_{0}^{2\pi} d \varphi \,  e^{i {\bm{\mu}} ( \eta \cdot \varepsilon_{(1)} \cos \varphi - \eta \cdot \varepsilon_{(1)} \sin \varphi)/( \eta \cdot p) }  \,
e^{i n \varphi} \, \Phi_n (p) \, ,
\ee
where we have used the relations (\ref{sol2s}), (\ref{sol4}) and introduced the notation for the polarization vectors
\be \lb{pv-11}
\begin{array}{c}
\varepsilon_{(1)} := \Lambda( A_{(p)}) \, \overset{_{\mathrm{\,o}}}{\varepsilon}_{(1)} \, ,\;\;\; \varepsilon_{(2)} := \Lambda( A_{(p)}) \, \overset{_{\mathrm{\,o}}}{\varepsilon}_{(2)} \, .
\end{array}
\ee
Let us apply the following substitution on the right-hand side (\ref{nrez-ra})
\be \lb{nrez-ra1}
\frac{\eta \cdot \varepsilon_{(1)}}{\sqrt{\overset{\phantom{}}{(\eta \cdot \varepsilon_{(1)})^2+(\eta \cdot \varepsilon_{(2)})^2}}} = \sin \alpha \, , \;\;\;
\frac{\eta \cdot \varepsilon_{(2)}}{\sqrt{\overset{\phantom{}}{(\eta \cdot \varepsilon_{(1)})^2+(\eta \cdot \varepsilon_{(2)})^2}}} = \cos \alpha \, ,
\ee
then one gets
\be \lb{nrez-ra2}
\Psi(p,\eta) = f(\eta \cdot \eta, \eta \cdot p) \sum_{n\in \mathbb{Z}} \int\limits_{0}^{2\pi} d \varphi \,  e^{- i {\bm{\mu}} \frac{\sqrt{\overset{\phantom{}}{(\eta \cdot \varepsilon_{(1)})^2+(\eta \cdot \varepsilon_{(2)})^2}}}{( \eta \cdot p)} \sin (\varphi-\alpha)+i n \varphi}  \, \Phi_n (p) \, ,
\ee
On the right-hand side of relation (\ref{nrez-ra2}), we change the integration variable ($\varphi^\prime = \varphi - \alpha$), use the integral representation of the Bessel functions $J_n(x)$ of integer order and perform the back substitution (\ref{nrez-ra1}). As a result, we obtain the following representation for the field $\Psi(p,\eta)$ as a series
\be \lb{nnrez1}
\Psi(p,\eta) = 2 \pi f(\eta \cdot \eta, \eta \cdot p)  \, \sum_{n \in \mathbb{Z}} J_n \!\! \left ( \bm{\mu} \frac{\sqrt{\overset{\phantom{}}{(\eta \cdot \varepsilon_{(1)})^2+(\eta \cdot \varepsilon_{(2)})^2}}}{\eta \cdot p} \right )\!
\exp \left \{i n \, \mathrm{arctg} \left ( \frac{\eta \cdot \varepsilon_{(1)}}{\eta \cdot \varepsilon_{(2)}} \right)\right \} \Phi_{n}(p) \,.
\ee

The resulting expression (\ref{nnrez1}) for a relativistic field of infinite spin uses the Bessel functions, which is natural for the massless
 $4D$ representations of the Poincar\'{e} group, because the irreducible representations of the small group $ISO(2)$
can be realized in the space the Bessel functions \cite{Vilen}.

\subsection{Singular solution}

As shown in Appendix 1, the solution to equations (\ref{fe2a}), (\ref{fe2b}), (\ref{fe2c}) for the kernel
at $\eta^+ = 0$ has the form
\be \lb{sol-2b}
\mathcal{A}(\overset{_{\mathrm{\;o}}}{p},\eta,\varphi) \ = \
\delta(\eta^+) \, \delta(\eta_1 \sin \varphi + \eta_{2} \cos \varphi)\,
e^{\,\frac{i}{2}\, {\bm{\rho}} \,\eta^-/\left(\eta_1 \cos \varphi - \eta_{2} \sin \varphi\right)  } \, f(\eta_1 \cos \varphi - \eta_{2} \sin \varphi) \, ,
\ee
where $\eta^\pm=\eta_0 \mp \eta_3$ and $f(x)$ is an arbitrary function.

After introducing the vectors (\ref{sol2s}) and the 4-vector $\overset{_{\mathrm{\,o}}}{\varepsilon}$ with the coordinates
\be \lb{sol2s0}
(\overset{_{\mathrm{\,o}}}{\varepsilon})_\nu = \Bigl(\frac{1}{2E},0,0,-\frac{1}{2E}\Bigr)\,,
\ee
expression (\ref{sol-2b}) becomes
\be \lb{sol-2c}
\mathcal{A}(\overset{_{\mathrm{\;o}}}{p},\eta,\varphi) \ = \
\delta(\eta\cdot \overset{_{\mathrm{\;o}}}{p}) \, \delta(\eta\cdot \overset{_{\mathrm{\,o}}}{\varepsilon}_{(2)}(\varphi))\,
e^{i {\bm{\mu}} \, \eta\cdot \overset{_{\mathrm{\,o}}}{\varepsilon}  /
(\eta\cdot \overset{_{\mathrm{\,o}}}{\varepsilon}_{(1)}(\varphi) )}  \,
f(\eta\cdot \overset{_{\mathrm{\,o}}}{\varepsilon}_{(1)}(\varphi)) \, .
\ee

According to (\ref{unpgr8}), the expression for the kernel $\mathcal{A}(p,\eta,\varphi)$ at arbitrary momentum is obtained from the expression (\ref{sol-2c}) by replacing $\overset{_{ \mathrm{\;o}}}{p} \to p$ and substitution
\be \lb{repl-2c}
\eta_\nu\ \ \to \ \ [\Lambda^{-1}(A_{(p)})\eta]_\nu \, .
\ee
This leads to the following expression:
\be \lb{sol-2d}
\mathcal{A}(p,\eta,\varphi) \ = \
\delta(\eta\cdot p) \, \delta(\eta\cdot {\varepsilon}_{(2)}(\varphi))\,
e^{i {\bm{\mu}} \, \eta\cdot {\varepsilon}  /
(\eta\cdot {\varepsilon}_{(1)}(\varphi) )}  \,
f(\eta\cdot {\varepsilon}_{(1)}(\varphi)) \, ,
\ee
where we have introduced the polarization vectors (\ref{sol4}) and the vector
\be \lb{sol9}
\varepsilon = \Lambda(A_{(p)}) \; \overset{_{\mathrm{\,o}}}{\varepsilon} \,  ,
\ee
obtained by transforming the vector (\ref{sol2s0}).
The vector (\ref{sol9}) is light-like and transverse to the vectors $\varepsilon_{(1)}(\varphi)$, $\varepsilon_{(2)}(\varphi)$:
\be \lb{sor1b}
\varepsilon \cdot \varepsilon  = 0\, , \qquad
\varepsilon\cdot \varepsilon_{(1)}(\varphi)  = \varepsilon\cdot \varepsilon_{(2)}(\varphi) = 0  \,.
\ee
Moreover, it obeys the condition
\be \lb{sor2b}
\varepsilon \cdot p  = 1\,.
\ee
Expression (\ref{sol-2d}) coincides with the generalized Wigner operator found in \cite{ShTor1}, \cite{ShTor2}.

Relation (\ref{unpgr4}), generates with the help of the generalized Wigner operator (\ref{sol-2d}), a Lorentz-covariant field $\Psi(p,\eta)$ of the following form:
\be \lb{nrez1-b}
\Psi(p,\eta) = \int\limits_{0}^{2\pi} d \varphi \,  \delta(\eta\cdot p) \, \delta(\eta\cdot {\varepsilon}_{(2)}(\varphi))\,
e^{i {\bm{\mu}} \, \eta\cdot {\varepsilon}  /
(\eta\cdot {\varepsilon}_{(1)}(\varphi) )}  \,
f(\eta\cdot {\varepsilon}_{(1)}(\varphi)) \, \Phi (p, \varphi) \, .
\ee
As shown in Appendix 2, the square of the Pauli-Lubanski vector takes the value $-{\bm{\mu}}^2$ on this field as well.
This means that the on the fields $\Psi(p, \eta)$, defined in (\ref{nrez1-b}), the irreducible representation
of the covering group $ISL(2,\mathbb{C})$ with infinite spin is realized.

Due to the presence of the $\delta$-function $\delta(\eta\cdot p)$ in expression (\ref{nrez1-b}), the equation of motion of the field $\Psi(p,\eta)$ is
\be \lb{eq-vec-1}
(\eta\cdot p)\,\Psi(p,\eta) = 0 \, .
\ee
Moreover, using (\ref{skaz-2}) it is easy to show that the field (\ref{nrez1-b}) obeys the equation
\be \lb{eq-vec-2}
\left[i\sqrt{-(\eta\cdot\eta)}\,(p\cdot\frac{\partial}{\partial\eta})+{\bm{\mu}}\right]\Psi(p,\eta) = 0 \, .
\ee
Additional condition $f(\eta\cdot {\varepsilon}_{(1)}(\varphi))=\delta(\eta\cdot {\varepsilon}_{(1)}(\varphi)-1)$ , fixing the function $f(\eta\cdot {\varepsilon}_{(1)}(\varphi))$ in (\ref{nrez1-b}), leads to the equation
\be \lb{eq-vec-3}
\big[(\eta\cdot\eta)+1\big]\,\Psi(p,\eta) = 0
\ee
for the field (\ref{nrez1-b}).
As a result, the equation (\ref{eq-vec-2}) becomes:
\be \lb{eq-vec-4}
\left[i\,(p\cdot\frac{\partial}{\partial\eta})+{\bm{\mu}}\right]\Psi(p,\eta) = 0 \, .
\ee
Together with the massless condition $p^2\,\Psi(p,\eta) = 0$, the equations (\ref{eq-vec-1}), (\ref{eq-vec-3}),
(\ref {eq-vec-4}) are the Bargmann-Wigner equations for infinite spin fields depending on an additional vector
variable $\eta$ \cite{Wig39,Wig48,BarWig48}.

\setcounter{equation}{0}
\section{Relativistic fields with an additional spinor variable}  \lb{sect-5}

Let us now consider the case when a commuting Weyl spinor is taken as an additional variable $y$
in the field description (\ref{unpgr4}), (\ref{unpgr4a}) of massless representations of infinite spin. Previously, such a description was used in \cite{BFIR}, \cite{BFI}.
In this case, the relativistic field $\Psi(p,u,\bar{u})\equiv \Psi(p_m,u^{\alpha},\bar{u}^{\dot\alpha})$ is a
function of the 4-momentum $p$, 2-component commuting Weyl spinor $u^{\alpha}$, $\alpha=1,2$, and its complex
conjugate spinor $\bar{u}^{\dot{\alpha} } = (u^{\alpha})^{*}$. Recall that the Lorentz transformation matrices
and relativistic $\sigma$ matrices have the following spinor indices $A_{\alpha}{}^{\beta}$,
$(A^\dagger)^{\dot\alpha}{ }_{\dot\beta}$, $(\sigma^n)_{\alpha\dot\beta}$ by definition. \footnote{Raising and lowering Weyl indices
is carried out using antisymmetric tensors
$\epsilon_{{\alpha} {\beta}}=
 -\epsilon^{{\alpha} {\beta}}$, $\epsilon_{\dot{\alpha}
 \dot{\beta}}=-\epsilon^{\dot{\alpha} \dot{\beta}}$,
 $\epsilon_{{1} {2}} = \epsilon_{\dot{1} \dot{2}} =1$
as follows: ${u}_{{\alpha}} = \epsilon_{{\alpha}
 {\beta}} u^{{\beta}}$, $\bar{u}_{\dot{\alpha}} =
 \epsilon_{\dot{\alpha} \dot{\beta}} \bar
 u^{\dot{\beta}}$.}

According to (\ref{unpgr4}), the field $\Psi(p,u,\bar{u})$ is found from the Wigner wave function
$\Phi(p,\varphi)$ in the form
\be \lb{dsp2}
\Psi(p,u,\bar{u}) \ = \ \int\limits_{0}^{2\pi} d \varphi \, \mathcal{A}(p,u,\bar{u},\varphi)\,
\Phi (p, \varphi)\,,
\ee
where the generalized Wigner operator $\mathcal{A}(p,u,\bar{u},\varphi)$ maps a function of $\varphi$
into a function depending on $u^{\alpha}$, $\bar{ u}^{\dot\alpha}$.

The Lorentz transformation of the field $\Psi(p,u,\bar{u})$ looks like (compare with the formula (\ref{unpgr41}) for fields with integer helicities)
\be \lb{dsp3}
\Psi^{\prime}(p, u, \bar{u}) \  =  \ [U(A) \Psi] (p,u,\bar{u})  \ =  \ \Psi(\Lambda^{-1} p, u A, A^{\dagger}\bar{u} ) \, ,
\ee
where $(u A)^\alpha=u^\beta A_\beta^{\;\; \alpha}$, $(A^{\dagger}\bar{u})^{\dot\alpha} = (A^\dagger)^{\dot\alpha}{}_{\dot\beta}
 \bar{u}^{\dot\beta}$ and the matrix $A \in SL(2,\mathbb{C})$.
The matrix $\Lambda \in SO^{\uparrow}(1,3)$ is related to $A$ by (\ref{Alambd}). Consider the function
(5.1) and discuss its possible structure. If this function is a polynomial in spinor variables, i.e.
$\Psi(p,u,\bar{u})=u^{\alpha}...\bar{u}^{\dot{\beta}}... \psi_{{\alpha}...\dot{\beta}...}(p)$ then we get that the relation
(5.1) reduces to standard transformation law for  finite-dimensional Lorentz group representation what does not corresponds to infinite
spin  representation of the Poincare group. Therefore, we must assume that the function
$\Psi(p,u,\bar{u})$ cannot be a polynomial in additional spinor variables. This means that it must
carry out the infinite-dimensional representation of the Lorentz group.

Let us find an explicit form of the kernel $\mathcal{A}(p,u,\bar{u},\varphi)$, which transforms the Wigner wave functions
  $\Phi(p,\varphi)$ into relativistic fields $\Psi(p,u,\bar{u})$.
The transformation law (\ref{unpgr3}) for $\Phi (p, \varphi)$ and the transformation law (\ref{dsp3}) for $\Psi(p, u, \bar{u})$ yield the equations defining the kernel $\mathcal{A}(p,u,\bar{u},\varphi)$:
\be \lb{dsp4}
\begin{array}{rcl}
\mathcal{A} (\Lambda^{-1} p, u A, A^{\dagger}\bar{u} , \varphi)  &=&  \ e^{- i \vec{b}_{A,\Lambda^{-1} p} \vec{t}_{\varphi+\theta_{A,\Lambda^{-1} p}}} \ \mathcal{A}(p,u,\bar{u},\varphi+\theta_{A,\Lambda^{-1} p}) \\ [6pt]
&=& \mathcal{A}(p,\eta,\varphi^\prime)\,\mathcal{D}_{\varphi^\prime\varphi}(\theta_{A,\Lambda^{-1} p},\vec{b}_{A,\Lambda^{-1} p})\, .
\end{array}
\ee
These equations are spinor analogues of equations (\ref{unpgr7}).

Similar to the additional vector variable case considered in the previous section (see (\ref{unpgr8}),
(\ref{unpgr9})), the equation (\ref{dsp4}) implies two important relations for the kernel
$\mathcal{A}(p,u,\bar{u},\varphi)$:
\begin{description}
\item[1)]
The kernel $\mathcal{A}(p,...)$ for arbitrary momentum $p$ is determined from the kernel $\mathcal{A}(\overset{_{\mathrm{\;o}}}{p},.. .)$ for the test momentum $\overset{_{\mathrm{\;o}}}{p}$ as follows:
\be \lb{dsp5}
\mathcal{A} (p, u, \bar{u}, \varphi) \ =  \ \mathcal{A}(\overset{_{\mathrm{\;o}}}{p} ,u A_{(p)},A_{(p)}^{\dagger}\bar{u} ,\varphi) \, .
\ee
\item[2)]
The action of the small group, when $A = h$, on the kernel $\mathcal{A}(\overset{_{\mathrm{\;o}}}{p},...)$ for the test momentum $\overset{_{\mathrm{\;o}}}{p}$ is written as
\be \lb{dsp6}
\mathcal{A}(\overset{_{\mathrm{\;o}}}{p},u h, h^{\dagger}\bar{u} , \varphi)  \ = \  e^{- i \vec{b} \cdot \vec{t}_{\varphi+\theta}} \mathcal{A}(\overset{_{\mathrm{\;o}}}{p},u,\bar{u},\varphi+\theta) \, .
\ee
\end{description}
In the left-hand side and right-hand side of relation (\ref{dsp6}), as in the vector case
(\ref{unpgr9}), the parametrization of the matrix $h \in ISO(2)$ defined in (\ref{fr8}) is used.

Let us find from (\ref{dsp6}) the kernel $\mathcal{A} (\overset{_{\mathrm{\;o}}}{p},u ,\bar{u},\varphi)$
at test momentum $ \overset{_{\mathrm{\;o}}}{p}$.

To do this, consider equality (\ref{dsp6}) in the infinitesimal case, which gives three differential
equations.
\begin{itemize}
\item In the case when $\theta$ is small and the vector $\vec{b}=0$, equation (\ref{dsp6}) becomes
\be \lb{seo1}
\frac{i}{2}  \Bigl ( u^1 \frac{\partial}{\partial u^1}  - u^2 \frac{\partial}{\partial u^2}  - \bar{u}^{\dot{1}} \frac{\partial}{\partial \bar{u}^{\dot{1}}}  + \bar{u}^{\dot{2}} \frac{\partial}{\partial \bar{u}^{\dot{2}}} \Bigr  ) \mathcal{A} (\overset{_{\mathrm{\;o}}}{p}, u, \bar{u}, \varphi)  = \frac{\partial}{\partial \varphi} \, \mathcal{A} (\overset{_{\mathrm{\;o}}}{p}, u, \bar{u}, \varphi)\,.
\ee
\item In the case when $\theta = 0$ and the vector $\vec{b}=(b_1,b_2)$ is small, the equation
(\ref{dsp6}) gives
\be \lb{seo4}
\left\{
\begin{array}{rcl}
\displaystyle
2i\,\bar{u}^{\dot{1}} \frac{\partial}{\partial \bar{u}^{\dot{2}}} \,  \mathcal{A} (\overset{_{\mathrm{\;o}}}{p}, u, \bar{u}, \varphi)
&=&
{\bm{\rho}} \, e^{i \varphi} \mathcal{A} (\overset{_{\mathrm{\;o}}}{p}, u, \bar{u}, \varphi)\,,
\\ [8pt]
\displaystyle
2i\,u^1 \frac{\partial}{\partial u^2} \,  \mathcal{A} (\overset{_{\mathrm{\;o}}}{p}, u, \bar{u}, \varphi)  &=&
{\bm{\rho}} \, e^{- i \varphi} \mathcal{A} (\overset{_{\mathrm{\;o}}}{p}, u, \bar{u}, \varphi)\,.
\end{array}
\right.
\ee
\end{itemize}
The general solution of the system of equations (\ref{seo4}) is written as follows:
\be \lb{seo6}
\mathcal{A} (\overset{_{\mathrm{\;o}}}{p}, u, \bar{u}, \varphi)  = \exp \left\{ -\frac{i}{2} \, {\bm{\rho}} \, \Bigl ( \frac{u^2}{u^1} \, e^{- i \varphi} + \frac{\bar{u}^{\dot{2}}}{\bar{u}^{\dot{1}}} \, e^{i \varphi} \Bigr ) \right\} f( u^1, \bar{u}^{\dot{1}}, \varphi)  \, ,
\ee
where $f(u^1, \bar{u}^{\dot{1}}, \varphi)$ is an arbitrary function of three variables $u^1$, $\bar{u}^{\dot{1}}$, $\varphi$.
Substituting (\ref{seo6}) into (\ref{seo1}), we get an equation for this function $f(u^1, \bar{u}^{\dot{1}}, \varphi)$:
\be \lb{seo7}
\frac{i}{2} \Bigl ( u^1 \frac{\partial}{\partial u^1} - \bar{u}^{\dot{1}} \frac{\partial}{\partial \bar{u}^{\dot{1}}} \Bigr ) f( u^1, \bar{u}^{\dot{1}}, \varphi) =
\frac{\partial}{\partial \varphi} \, f(u^1, \bar{u}^{\dot{1}}, \varphi) \, .
\ee
It follows from equation (\ref{seo7}) that the function $f(u^1, \bar{u}^{\dot{1}}, \varphi)$ does not
depend on the phase variable $[2\arg(u^1)-\varphi]$ but depends only on the complex number
$u^1 e^{\frac{i}{2} \varphi}$ and its conjugate $\bar{u}^{\dot{1} } e^{ - \frac{i}{2} \varphi}$. Thus,
the kernel $\mathcal{A} (\overset{_{\mathrm{\;o}}}{p}, u, \bar{u}, \varphi)$ in the general case has the
form:
\be \lb{seo8}
\mathcal{A} (\overset{_{\mathrm{\;o}}}{p}, u, \bar{u}, \varphi) =  \exp \left\{ -\frac{i}{2}  \, {\bm{\rho}} \, \Bigl ( \frac{u^2}{u^1} e^{- i \varphi} + \frac{\bar{u}^{\dot{2}}}{\bar{u}^{\dot{1}}} e^{i \varphi} \Bigr ) \right\} f( u^1 e^{\frac{i}{2} \varphi}, \bar{u}^{\dot{1}} e^{ - \frac{i}{2} \varphi})\,.
\ee

The kernel $\mathcal{A}(p, u, \bar{u}, \varphi)$ for arbitrary momentum $p$ is determined from the
kernel $\mathcal{A} (\overset{_{\mathrm{\;o}}}{p}, u, \bar{u}, \varphi)$ for the test
momentum $\overset{_{\mathrm{\;o}}}{p}$ with the help of relation (\ref{dsp5}). That is, if we make
substitution
\be \lb{repl}
u^{\alpha} \ \to \ [u A_{(p)}]^{\alpha}\,,\qquad \bar{u}^{\dot\alpha} \ \to \ [A_{(p)}^{\dagger}\bar{u}]^{\dot\alpha}\,,
\ee
in relation (\ref{seo8}), we get an expression for $\mathcal{A}(p, u, \bar{u}, \varphi)$:
\be \lb{seo8-f}
\mathcal{A} (p, u, \bar{u}, \varphi) =  \exp \left\{ -\frac{i}{2}  \, {\bm{\rho}} \, \left(
\frac{[u A_{(p)}]^2}{[u A_{(p)}]^1}\, e^{- i \varphi} +
\frac{[A_{(p)}^{\dagger}\bar{u}]^{\dot{2}}}{[A_{(p)}^{\dagger}\bar{u}]^{\dot{1}}}\, e^{i \varphi} \right) \right\}
f( [u A_{(p)}]^1 e^{\frac{i}{2} \varphi}, [A_{(p)}^{\dagger}\bar{u}]^{\dot{1}} e^{ - \frac{i}{2} \varphi})\,.
\ee
As a result of this relation and the definition (\ref{dsp2}), the relativistic field $\Psi(p,u,\bar{u})$ of a massless particle of infinite spin has the form
\be \lb{dsp2-sp-f}
\Psi(p,u,\bar{u})  =  \int\limits_{0}^{2\pi} d \varphi \, \exp \left\{ -\frac{i}{2}  \, {\bm{\rho}} \, \left(
\frac{[u A_{(p)}]^2}{[u A_{(p)}]^1}\, e^{- i \varphi} +
\frac{[A_{(p)}^{\dagger}\bar{u}]^{\dot{2}}}{[A_{(p)}^{\dagger}\bar{u}]^{\dot{1}}}\, e^{i \varphi} \right) \right\}
f( [u A_{(p)}]^1 e^{\frac{i}{2} \varphi}, [A_{(p)}^{\dagger}\bar{u}]^{\dot{1}} e^{ - \frac{i}{2} \varphi})\, \Phi (p, \varphi)\,.
\ee

Let us find the value of the Casimir operator $\hat{W}^2$ under its action on the field
$\Psi(p,u,\bar{u})$. To do this, as in the case of fields with an additional vector variable, it suffices to calculate the value of $\hat{W}^2$ on the kernel $\mathcal{A} (p, u, \bar{u}, \varphi)$, which is found in (\ref{seo8-f}). The Casimir operator $\hat{W}^2$ written in terms of generators acting on the field $\Psi(p,u,\bar{u})$ and depending on the spinor variables $u^{\alpha}$, $ \bar u^{\dot\alpha}$ and 4-momentum $p^m$, has the form \cite{BuchIFKr}
\be \lb{seo13a0}
\hat{W}^2 = (u^{\alpha} p_{\alpha\dot{\alpha}} \bar{u}^{\dot{\alpha}} ) \Bigl( \frac{\partial}{\partial \bar{u}^{\dot{\beta}}}
p^{\dot{\beta}\beta} \frac{\partial}{\partial u^{\beta}} \Bigr )\,,
\ee
where $p_{\alpha\dot{\alpha}}=p_m(\sigma^m)_{\alpha\dot{\alpha}}$ and $p^{\dot{\beta}\beta}=
\epsilon^{\beta\alpha}\epsilon^{\dot\beta\dot\alpha}
p_{\alpha\dot{\alpha}}$.

The action of this operator on the kernel (\ref{seo8-f}) is easily calculated and taking into account the equalities
$p^{\dot{\alpha}\beta}(A_{(p)})_{\beta}{}^{1}=(A^\dagger_{(p)})^{\dot 1}{}_{\dot\beta}p^{\dot{\beta}\alpha}=0$,
$(A^\dagger_{(p)})^{\dot 1}{}_{\dot\alpha}p^{\dot{\alpha}\beta}(A_{(p)})_{\beta}{}^{1}=
(A^\dagger_{(p)})^{\dot 1}{}_{\dot\alpha}p^{\dot{\alpha}\beta}(A_{(p)})_{\beta}{}^{2}=
(A^\dagger_{(p)})^{\dot 2}{}_{\dot\alpha}p^{\dot{\alpha}\beta}$
\newline
$(A_{(p)})_{\beta}{}^{1}=0, \, (A^\dagger_{(p)})^{\dot 2}{}_{\dot\alpha}p^{\dot{\alpha}\beta}(A_{(p)})_{\beta}{}^{2}=2E$ leads to
\be \lb{seo14}
\hat{W}^2 \, \Psi(p,u,\bar{u}) = -{\bm{\mu}}^2 \,  \Psi(p,u,\bar{u}) \, ,
\ee
where the dimensional constant ${\bm{\mu}}$, as in the case of the vector variables $y=\eta$, is equal
to ${\bm{\mu}}=E{\bm{\rho}}$ (see .(\ref{mu})). Thus, the obtained fields $\Psi(p,u,\bar{u})$ satisfy
the necessary condition that these fields form the space of an irreducible massless infinite spin
representation of the group $ISL(2,\mathbb{C})$.

Equation (\ref{seo14}) can be thought of as the equation of motion for the field (\ref{dsp2-sp}).
However, as follows from (\ref{seo13a0}), the operator $\hat{W}^2$ is the product of two scalar operators
$(u^{\alpha} p_{\alpha\dot{\alpha}} \bar{u}^{\dot{\alpha}} )$ and
$\displaystyle\Bigl( \frac{\partial}{\partial \bar{u}^{\dot{\beta}}}
p^{\dot{\beta}\beta} \frac{\partial}{\partial u^{\beta}} \Bigr )$.
Therefore, in the papers \cite{BuchKrTak,BuchIFKr}, instead of equation (\ref{seo14}),
more stronger conditions were imposed on the field $\Psi(p,u,\bar{u})$ depending on additional spinor
variables
\begin{eqnarray}
\lb{eq-sp-1}
\left(u^{\alpha} p_{\alpha\dot{\alpha}} \bar{u}^{\dot{\alpha}}-{\bm{\mu}} \right)\Psi(p,u,\bar{u}) &=&0\,, \\
\lb{eq-sp-2}
\left( \frac{\partial}{\partial \bar{u}^{\dot{\beta}}}
p^{\dot{\beta}\beta} \frac{\partial}{\partial u^{\beta}} +{\bm{\mu}} \right)\Psi(p,u,\bar{u})&=&0\,.
\end{eqnarray}
Note that the kernel (\ref{seo8-f}) does not satisfy equations (\ref{eq-sp-1}), (\ref{eq-sp-2}).
To fulfill these equations, a kernel  in \cite{BuchKrTak,BuchIFKr} was taken under assumption that
dependence on $u^{\alpha} p_{\alpha\dot{\alpha}} \bar{u}^{\dot{\alpha}}$ is fixed due to
the presence of an additional function
$\delta(u^{\alpha} p_{\alpha\dot{\alpha}} \bar{u}^{\dot{\alpha}}-{\bm{\mu}})$ in solution (\ref{seo8-f}).

When describing representations of the Poincar\'{e} group in the space with an additional spinor
variable, it is convenient to use the finite-dimensional Wigner operator $A_{(p)}$ written in terms
of twistor spinors (in the case of massive representations $ISL(2,\mathbb{C})$, such a parametrization
was used in \cite{IsPod0})
\be \lb{seo9}
A_{(p)} =
\left(
\begin{array}{cc}
\displaystyle\frac{1}{\sqrt{2E}}\;\pi_1 \ & \sqrt{2E}\,\lambda_1 \\ [8pt]
\displaystyle\frac{1}{\sqrt{2E}}\;\pi_2 \ & \sqrt{2E}\,\lambda_2 \\
\end{array}
\right)
\, ,
\ee
where $\pi_{\alpha}$ and $\lambda_{\alpha}$ $(\alpha=1,2)$ are the commuting Weyl spinors. Then, an arbitrary massless 4-momentum $p$ is represented by the Cartan-Penrose relation
\be \lb{seo12}
(p \,\sigma)_{\alpha\dot\beta} \equiv p_{\alpha\dot\beta}=
\bigl( A_{(p)} \, (\overset{_{\mathrm{\;o}}}{p} \sigma)
\, A_{(p)}^\dagger \bigr)_{\alpha\dot\beta}
= \pi_{\alpha} \bar{\pi}_{\dot{\beta}} \, .
\ee
The $\pi_{\alpha}$ spinor defines half of the components of the Penrose twistor. The second spinor $\lambda_{\alpha}$, not included in relation (\ref{seo12}), obeys the only (complex) equation
\be \lb{seo11}
\pi^{\alpha} \lambda_{\alpha} = 1 \,,
\ee
which follows from the condition $\det A_{(p)}=1$ for the matrix $A_{(p)}\in SL(2,\mathbb{C})/ISO(2)$.
The components of the spinor $\lambda$ are functions of the components of the spinor $\pi$: $\lambda_{\alpha}=\lambda_{\alpha}(\pi)$. As $\lambda$ we can take the spinor
\be \lb{la-sol}
\lambda_{\alpha} = \frac{(\sigma_0)_{\alpha\dot\alpha}\bar\pi^{\dot\alpha}}{\pi^\beta(\sigma_0)_{\beta\dot\beta}\bar\pi^{\dot\beta}} \,,
\ee
obtained in an $SL(2,\mathbb{C})$-noncovariant way from the spinor $\pi$.

Now, using the explicit form of the Wigner operator (\ref{seo9}), we get the following expression for the kernel (\ref{seo8-f}):
\be \lb{seo13}
\mathcal{A} (\pi,\bar\pi, u, \bar{u}, \varphi) = \exp \left\{-i  \, {\bm{\mu}} \,
\Bigl ( \frac{ u^{\alpha} \lambda_{\alpha}}{u^{\beta} \pi_{\beta}} \, e^{- i \varphi} +
\frac{ \bar{u}^{\dot{\alpha}} \bar{\lambda}_{\dot{\alpha}}}{\bar{u}^{\dot{\beta}} \bar{\pi}_{\dot{\beta}}} \, e^{ i \varphi} \Bigr ) \right\}
f(u^{\gamma} \pi_{\gamma} \, e^{\frac{i}{2} \varphi}, \bar{u}^{\dot{\gamma}} \bar{\pi}_{\dot{\gamma}} \, e^{-\frac{i}{2} \varphi} )\,,
\ee
where $f$ is an arbitrary function.
Using expressions (\ref{seo13}) for the kernel $\mathcal{A}(p,u,\bar{u},\varphi)$ and formula (\ref{dsp2}), we find a twistor field $\Psi(\pi,\bar\pi,u,\bar{u})$ of a particle of infinite spin:
\be \lb{dsp2-sp}
\Psi(\pi,\bar\pi,u,\bar{u})  =  \int\limits_{0}^{2\pi} d \varphi \, \exp \left\{-i  \, {\bm{\mu}} \,
\Bigl ( \frac{ u^{\alpha} \lambda_{\alpha}}{u^{\beta} \pi_{\beta}} \, e^{- i \varphi} +
\frac{ \bar{u}^{\dot{\alpha}} \bar{\lambda}_{\dot{\alpha}}}{\bar{u}^{\dot{\beta}} \bar{\pi}_{\dot{\beta}}} \, e^{ i \varphi} \Bigr ) \right\}
f(u^{\gamma} \pi_{\gamma} \, e^{\frac{i}{2} \varphi}, \bar{u}^{\dot{\gamma}} \bar{\pi}_{\dot{\gamma}} \, e^{-\frac{i}{2} \varphi} )\, \Phi (p, \varphi)\,,
\ee
which corresponds to the field (\ref{dsp2-sp-f}) but depends on the twistor spinor $\pi_\alpha$.

In the case of the twistor representation (\ref{seo12}) for the 4-momentum $p^m$, the operator (\ref{seo13a0}) is equal to
\be \lb{seo13a}
\hat{W}^2 = (u^{\alpha} \pi_{\alpha} \bar\pi_{\dot{\alpha}} \bar{u}^{\dot{\alpha}} ) \Bigl( \frac{\partial}{\partial \bar{u}^{\dot{\beta}}} \bar\pi^{\dot{\beta}} \pi^{\beta} \frac{\partial}{\partial u^{\beta}} \Bigr )
\ee
and, like (\ref{seo14}), we have:
\be \lb{seo14-tw}
\hat{W}^2 \, \Psi(\pi,\bar\pi,u,\bar{u}) = -{\bm{\mu}}^2 \,  \Psi(\pi,\bar\pi,u,\bar{u}) \, .
\ee

A twistor formulation of a massless particle of infinite spin in four-dimensional space-time was
constructed in \cite{BFIR,BFI}, where a twistor equation similar to (\ref{eq-sp-1}) was postulated
\be \lb{eq-tw-3}
\left(\pi^{\alpha}u_{\alpha} \bar{u}_{\dot{\alpha}}\bar\pi^{\dot{\alpha}} - {\bm{\mu}} \right)\Psi(\pi,\bar\pi,u,\bar{u}) =0
\ee
or stronger equations were postulated
\begin{eqnarray}
\lb{eq-tw-4}
\left( \pi^{\alpha}u_{\alpha}  -\sqrt{{\bm{\mu}}} \right)\Psi(\pi,\bar\pi,u,\bar{u}) &=&0\,, \\ [6pt]
\lb{eq-tw-5}
\left( \bar{u}_{\dot{\alpha}}\bar\pi^{\dot{\alpha}} -\sqrt{{\bm{\mu}}} \right)\Psi(\pi,\bar\pi,u,\bar{u})&=&0\,.
\end{eqnarray}
Equations (\ref{eq-tw-3}) and (\ref{eq-tw-4})-(\ref{eq-tw-5}) fix the eigenvalue of part of the operator $\hat{W}^2$ in (\ref{seo13a}).
The eigenvalues of the second factor in $\hat{W}^2$ in (\ref{seo13a}) were given in \cite{BFIR,BFI} by the equations of motion of the twistor field
\begin{eqnarray}
\lb{eq-tw-1}
\left( \pi^{\beta}\frac{\partial}{\partial u^\beta} +i\sqrt{{\bm{\mu}}} \right)\Psi(\pi,\bar\pi,u,\bar{u}) &=&0\,, \\
\lb{eq-tw-2}
\left( \bar\pi^{\dot\beta}\frac{\partial}{\partial \bar{u}^{\dot{\beta}}} +i\sqrt{{\bm{\mu}}} \right)\Psi(\pi,\bar\pi,u,\bar{u})&=&0\,.
\end{eqnarray}
Equations (\ref{eq-tw-4}), (\ref{eq-tw-5}) (or (\ref{eq-tw-3})) and (\ref{eq-tw-1}), (\ref{eq-tw-2}) lead to the fulfillment of the irreducibility conditions (\ref{seo14-tw}) for the twistor field of infinite spin, which is a certain special case of the field (\ref{dsp2-sp}) and has the form
\be \lb{dsp2-sp-tw}
\Psi(\pi,\bar\pi,u,\bar{u}) =  \delta(\pi^{\alpha}u_{\alpha}  -\sqrt{{\bm{\mu}}})\,
\delta(\bar{u}_{\dot{\alpha}}\bar\pi^{\dot{\alpha}} -\sqrt{{\bm{\mu}}})
\int\limits_{0}^{2\pi} d \varphi \, e^{\displaystyle i  \, \sqrt{{\bm{\mu}}} \,
\bigl ( u^{\alpha} \lambda_{\alpha} \, e^{- i \varphi} +
\bar{u}^{\dot{\alpha}} \bar{\lambda}_{\dot{\alpha}} \, e^{ i \varphi} \bigr ) }
\, \Phi (p, \varphi)\,,
\ee
where we have redefined the function $\Phi$ by the function $f$: $f \Phi \to \Phi$ and the spinor $\lambda$ is defined in (\ref{la-sol}).

To remove the integration over $\varphi$ in the definition (\ref{dsp2-sp-tw}) of the field $\Psi(\pi,\bar\pi,u,\bar{u})$, we can use the following identity
\be \lb{bes-e1}
 u^{\alpha} \lambda_{\alpha} \, e^{- i \varphi} + \bar{u}^{\dot{\alpha}} \bar{\lambda}_{\dot{\alpha}} \, e^{ i \varphi} =
2\left ( \mathbb{R}e\{u^{\alpha} \lambda_{\alpha}\} \cos \varphi + \mathbb{I}m\{u^{\alpha} \lambda_{\alpha}\} \sin \varphi \right)
\ee
and apply the same reasoning as when deriving formula (\ref{nnrez1}). As a result, the twistor field of infinite spin is represented as expansion
\be \lb{bes-e2}
\Psi(\pi,\bar\pi,u,\bar{u}) = 2 \pi \delta(\pi^{\alpha}u_{\alpha}  -\sqrt{{\bm{\mu}}}) \, \delta(\bar{u}_{\dot{\alpha}}\bar\pi^{\dot{\alpha}} -\sqrt{{\bm{\mu}}})
\sum_{n \in \mathbb{Z}} J_n \left (-2 \sqrt{{\bm{\mu}}} \, |u^{\alpha} \lambda_{\alpha}| \right )
e^{ - i n \, \mathrm{arctg} \left( \frac{\mathbb{R}e\{u^{\alpha} \lambda_{\alpha}\}}{ \mathbb{I}m\{u^{\alpha} \lambda_{\alpha}\}} \right )} \Phi_n(p)
\ee
by the Bessel functions $J_n$.

Note that representations of infinite spin necessarily require a bitwistor description \cite{BFIR,BFI}, where the spinor $\pi_\alpha$ is included in the definition of one twistor, and the spinor component of the second twistor (in \cite{BFIR,BFI} it was denoted by $\rho_\alpha$), which has the dimension of the square root of the mass, was present in constraints identical to equations (\ref{eq-tw-3}), (\ref{eq-tw-4}), (\ref{eq-tw-5}), (\ref{eq-tw-1}), (\ref{eq-tw-2}). Thus, the spinor $u^{\alpha}$ used here multiplied by $\sqrt{{\bm{\mu}}}$ plays the role of the spinor component of the second twistor used in the bitwistor formulation of the particle of infinite spin \cite{BFIR,BFI}.

\setcounter{equation}{0}
\section{Conclusion} \lb{sect-6}

In this paper, we have introduced and studied the relativistic fields $\Psi(p, \eta)$ and
$\Psi(p, u, \bar{u})$, which describe the infinite spin irreducible unitary representations of
the Poincar\'{e} group in $4D$ Minkowski space. The resulting fields $\Psi(p, \eta)$ and $\Psi(p, u, \bar{u})$ are
defined on spaces parameterized by the 4-momentum $p$ and the additional variables: the commuting
4-vector $\eta ^\mu$ or commuting Weyl spinor $u^\alpha$, respectively. Such relativistic fields
are determined by the integral transformation of the Wigner function $\Phi(p, \varphi)$, on which the
unitary infinite-dimensional representation of the small group $ISO(2)$ of the massless 4-momentum is
realized. The transition from the functions $\Phi(p, \varphi)$ to the fields $\Psi(p, \eta)$ and
$\Psi(p, u, \bar{u})$ is carried out using the generalized Wigner operators
$ \mathcal{A}(p,\eta,\varphi)$ and $\mathcal{A} (p, u, \bar{u}, \varphi)$, respectively.
These operators are solutions to differential equations that arise due to infinitesimal
transformations of fields and the Wigner wave function. We showed that on the fields
$\Psi(p, \eta)$ and $\Psi(p, u, \bar{u})$ the Casimir operators of the Poincar\'{e} algebra
take values corresponding to massless infinite spin particles in the four-dimensional Minkowski space.

Note that expressions (\ref{sol3}), (\ref{sol-2d}) for the generalized Wigner operators
$\mathcal{A}(p,\eta,\varphi)$ and (\ref{seo8-f}) for $\mathcal{A} (p, u, \bar{u}, \varphi)$ contain
an arbitrary function $f$. After fixing it, the field $\Psi(p, \eta)$ obeys the well-known
Bargmann-Wigner equations \cite{Wig48,BarWig48}, while the field $\Psi(p, u, \bar{u})$ obeys
the equations from \cite{BFIR,BFI}. The constructions of fields of infinite spin proposed here,
in particular, the expression (\ref{dsp2-sp}) for the field $\Psi(p, u, \bar{u})$, can be useful
for the Lagrangian description of such fields, including the description of their interactions (see,
for example, \cite{ShTor1,ShTor2,ShTorZh}).

In this paper, we considered only massless fields describing representations of the Poincar\'{e}
group of infinite spin. However, the approach, which uses additional commuting variables,
also makes it possible to find the corresponding massless fields of finite spin (helicity).
By defining the generalized Wigner operators $\mathcal{A}(p,\eta,\varphi)$ and
$\mathcal{A} (p, u, \bar{u}, \varphi)$ from the equations that are the limit ${ \bm{\rho}}\to0$
equations (\ref{fe2a}), (\ref{fe2b}), (\ref{fe2c}) and (\ref{seo1}), (\ref{seo4}) , we found
the fields $\Psi(p, \eta)$ and $\Psi(p, u, \bar{u})$ describing massless states of fixed helicity.
In the standard approach, such particles are described by spin-tensor fields, which are either field
strengths or potentials. In the approach under consideration, such spin-tensor fields arise in the
expansion of $\Psi(p, \eta)$ and $\Psi(p, u, \bar{u})$ in additional variables $\eta$ or $u$. The
discussion in \cite{ZFed} shows that in the case of an additional spinor variable, the field
$\Psi(p, u, \bar{u})$ reproduces massless field strengths. However in the case of an additional
vector variable, the field $\Psi(p, \eta)$ reproduces the description of massless states in terms
of potentials with gauge symmetry. We plan to study this issue in more detail in the forthcoming papers.

\smallskip
\section*{Acknowledgment}
The work of I.L.B. and S.A.F. was supported by RSF grant No.\,21-12-00129.

\section*{Appendix\,1: \\
Solving equations for the kernel $\mathcal{A}(\overset{_{\mathrm{\;o}}}{p},\eta,\varphi)$ }
\def\theequation{A.\arabic{equation}}
\setcounter{equation}0

To find a solution to equation (\ref{fe2a}), it is convenient to introduce new variables\footnote{We do not change the arguments in
$\mathcal{A}(\overset{_{\mathrm{o}}}{p},\eta,\varphi)$ with each transformation of variables, implying that the old variables $\eta^m$ and $\varphi$ are expressed in terms of the new ones through these transformations.}
\be \lb{nev1}
\zeta = \eta^2 + i \eta^1 \, , \;\;\; \bar{\zeta} = \eta^2 - i \eta^1
\;\;\;\; \Leftrightarrow \;\;\;\;
\eta^1 =\frac{i}{2}(\bar{\zeta} - \zeta ) \; , \;\;\;
\eta^2 =\frac{1}{2}(\bar{\zeta} + \zeta )  \,,
\ee
in which equation (\ref{fe2a}) becomes
\be \lb{nev2}
- i \left( \zeta \frac{\partial}{\partial \zeta} - \bar{\zeta} \frac{\partial}{\partial \bar{\zeta}} \right)
 \mathcal{A}(\overset{_{\mathrm{\;o}}}{p},\eta,\varphi) \ = \ \frac{\partial}{\partial \varphi} \, \mathcal{A}(\overset{_{\mathrm{\;o}}}{p},\eta,\varphi)\,.
\ee
After introducing polar coordinates $\varrho\in \mathbb{R}$, $\alpha \in [0,2\pi]$ for
$\zeta$ and $\bar{\zeta}$:
\be \lb{nev2a}
\zeta = \varrho \, e^{-i \alpha} \, , \quad \bar{\zeta} = \varrho \, e^{i \alpha} \, ,
\ee
the equation (\ref{nev2}) becomes
\be \lb{nev3}
\left( \frac{\partial}{\partial \alpha} - \frac{\partial}{\partial \varphi} \right) \mathcal{A}(\overset{_{\mathrm{\;o}}}{p},\eta,\varphi) \ = \ 0 \, .
\ee
Equation (\ref{nev3}) shows that the kernel $\mathcal{A}(\overset{_{\mathrm{\;o}}}{p},\dots)$ does not
depend on $\gamma^{- } = \alpha - \varphi$ and depends only on the variable
$\gamma^{+} = \alpha + \varphi$. The latter is conveniently taken into account by introducing the
variables
\be \lb{nev4}
z := \zeta e^{- i \varphi} = \varrho e^{-i \gamma^{+}} \, , \;\;\; \bar{z} := \bar{\zeta} e^{i \varphi} = \varrho e^{i \gamma^{+}}  \, .
\ee
Thus, the kernel $\mathcal{A}(\overset{_{\mathrm{\;o}}}{p},\dots)$ that satisfies (\ref{fe2a}) must
have the following functional dependence:
\be \lb{anz1}
\mathcal{A}(\overset{_{\mathrm{o}}}{p},z,\bar{z},\eta^0, \eta^3) \, .
\ee

Let us move on to solving equations (\ref{fe2b}), (\ref{fe2c}). To do this, we introduce the light cone variables
\be \lb{nv1}
\eta^{\pm} := \eta^0 \pm \eta^3 \,.
\ee
Then the equations (\ref{fe2b}) and (\ref{fe2c}) are written as
\begin{eqnarray} \lb{ne1}
i\left(\eta^1 \frac{\partial}{\partial \eta^{-}} +\frac{1}{2}\, \eta^{+} \frac{\partial}{\partial \eta^1} \right) \mathcal{A}(\overset{_{\mathrm{\;o}}}{p},\eta,\varphi) &=&
 \frac{\bm{\rho}}{2} \, \cos\varphi \, \mathcal{A}(\overset{_{\mathrm{\;o}}}{p},\eta,\varphi)  \, ,
\\ [6pt]
\lb{ne2}
\quad\left(\eta^2 \frac{\partial}{\partial \eta^-} + \frac{1}{2}\, \eta^+ \frac{\partial}{\partial \eta^{2}} \right) \mathcal{A}(\overset{_{\mathrm{\;o}}}{p},\eta,\varphi) &=&
i \, \frac{\bm{\rho}}{2} \, \sin\varphi \,\mathcal{A}(\overset{_{\mathrm{\;o}}}{p},\eta,\varphi)  \, .
\end{eqnarray}
The sum and difference of equations (\ref{ne1}), (\ref{ne2}) leads to
\begin{eqnarray}
\lb{ne3}
 \left(z \frac{\partial}{\partial \eta^{-}} + \eta^{+} \frac{\partial}{\partial \bar{z} } \right) \mathcal{A}(\overset{_{\mathrm{\;o}}}{p},z,\bar{z},\eta^{\pm})&=&
\frac{\bm{\rho}}{2} \, \mathcal{A}(\overset{_{\mathrm{\;o}}}{p},z,\bar{z},\eta^{\pm})  \, ,
\\ [6pt]
\lb{ne4}
- \left(\bar{z} \frac{\partial}{\partial \eta^{-}} + \eta^{+} \frac{\partial}{\partial z } \right)\mathcal{A}(\overset{_{\mathrm{\;o}}}{p},z,\bar{z},\eta^{\pm}) &=&
\frac{\bm{\rho}}{2} \, \mathcal{A}(\overset{_{\mathrm{\;o}}}{p},z,\bar{z},\eta^{\pm}) \, ,
\end{eqnarray}
where variables (\ref{nev4}) have been used.

First, consider the case $\eta^+\neq 0$.

Multiply (\ref{ne3}) by $\bar{z}$, (\ref{ne4}) by $z$ and add them. As a result, we obtain the equation
\be \lb{ne5}
\left( \bar{z} \frac{\partial}{\partial \bar{z}} - z \frac{\partial}{\partial z} \right) \mathcal{A}(\overset{_{\mathrm{\;o}}}{p},z,\bar{z},\eta^{\pm}) \ = \
\bm{\rho} \, \frac{(z+\bar{z})}{2 \eta^+} \, \mathcal{A}(\overset{_{\mathrm{\;o}}}{p},z,\bar{z},
\eta^{\pm})\,.
\ee
Solution to this equation has the form
\be \lb{nsol1}
\mathcal{A}(\overset{_{\mathrm{\;o}}}{p},\eta,\varphi) = f (\eta^{\pm}, z \bar{z}) \, e^{\bm{\rho} (\bar{z} - z)/(2\eta^+)} \, ,
\ee
where $f(\eta^{\pm}, z \bar{z})$ is an arbitrary function of $\eta^{+}$, $\eta^{-}$ and $z \bar{z }$. Note that the solution (\ref{nsol1}) is easily found after passing from $z, \bar{z}$ to the variables $\varrho, \gamma^{+}$, according to (\ref{nev4}).

Consider now the difference between equation (\ref{ne3}) multiplied by $\bar{z}$ and equation (\ref{ne4}) multiplied by $z$. As a result, we obtain the equation
\be \lb{ne6}
\left[ 2 z \bar{z}  \frac{\partial}{\partial \eta^{-}} + \eta^{+} (\bar{z} \frac{\partial}{\partial \bar{z}} + z \frac{\partial}{\partial z} ) \right] \mathcal{A}(\overset{_{\mathrm{\;o}}}{p},z,\bar{z},\eta^{\pm}) = \frac{\bm{\rho}}{2}(\bar{z}-z) \mathcal{A}(\overset{_{\mathrm{\;o}}}{p},z,\bar{z},\eta^{\pm}) \, .
\ee
Now substituting the expression (\ref{nsol1}) into (\ref{ne6}), we find a new equation for the function $f (\eta^{\pm}, z \bar{z})$:
\be \lb{ne7}
\left[ 2 z \bar{z} \frac{\partial}{\partial \eta^{-}} + \eta^{+} (\bar{z} \frac{\partial}{\partial \bar{z}} + z \frac{\partial}{\partial z} ) \right] f (\eta^{\pm}, z \bar{z}) = 0 \, .
\ee
Since we consider the case $\eta^+ \neq 0$, the solution to equation (\ref{ne7}) is an arbitrary function \footnote{The function $f (\eta^{\pm}, z \bar{z})$ depends on three variables: $\eta^{\pm}$ and $\varrho$. Passing from these variables to $\eta^{+}$, $y_1 = \eta^{+} \eta^{-} - \varrho^2$, $y_2 = \eta^{+} \eta^{- } + \varrho^2$, we find that equation (\ref{ne7}) makes the function $f (\eta^{\pm}, z \bar{z})$ independent of $y_2$.}
\be \lb{psol}
f (\eta^{+} \eta^{-} - z \bar{z},\eta^{+})
\ee
of two variables $(\eta^{+} \eta^{-} - z \bar{z})$ and $\eta^{+}$.
As a result, in terms of the variables $z,\bar{z}, \eta^{\pm}$, the general solution of equations (\ref{fe2a}), (\ref{fe2b}), (\ref{fe2c})
has the following form:
\be \lb{gsol}
 \mathcal{A}(\overset{_{\mathrm{o}}}{p},z,\bar{z},\eta^{\pm}) = f (\eta^{+} \eta^{-} - z \bar{z},\eta^{+}) \, e^{\,\bm{\rho} (\bar{z} - z)/(2\eta^+)} \, .
\ee
After taking into account the equality $\eta^{+} \eta^{-} - z \bar{z}=\eta^m \eta_m:=\eta\cdot\eta$ and restoring the original variables, solution (\ref{gsol}) is written in the form (\ref{sol}).

Let us now consider the case $\eta^+ = 0$ and find a solution to equations (\ref{ne3}) and (\ref{ne4}). The condition $\eta^+ = 0$ can be explicitly taken into account in (\ref{ne3}) and (\ref{ne4}) by requiring that the kernel $\mathcal{A}$ is proportional to the corresponding $\delta$-function $ \delta(\eta^+)$. That is, in this case
\be \lb{sol-2}
\mathcal{A}(\overset{_{\mathrm{\;o}}}{p},z,\bar{z},\eta^{\pm})=\delta(\eta^+) \, \tilde{\mathcal{A}}(\overset{_{\mathrm{\;o}}}{p},z,\bar{z},\eta^{-})\,,
\ee
and equations (\ref{ne3}) and (\ref{ne4}) take the form
\begin{eqnarray}
\lb{eq-01}
z \frac{\partial}{\partial \eta^{-}} \, \tilde{\mathcal{A}}(\overset{_{\mathrm{\;o}}}{p},z,\bar{z},\eta^{-})&=&
\frac{\bm{\rho}}{2} \, \tilde{\mathcal{A}}(\overset{_{\mathrm{\;o}}}{p},z,\bar{z},\eta^{-})  \, ,
\\ [6pt]
\lb{eq-02}
- \bar{z} \frac{\partial}{\partial \eta^{-}} \, \tilde{\mathcal{A}}(\overset{_{\mathrm{\;o}}}{p},z,\bar{z},\eta^{-}) &=&
\frac{\bm{\rho}}{2} \, \tilde{\mathcal{A}}(\overset{_{\mathrm{\;o}}}{p},z,\bar{z},\eta^{-}) \, .
\end{eqnarray}
The consequence of these equations is the condition
\be \lb{sol-zz}
\left( z+\bar z\right)\tilde{\mathcal{A}}(\overset{_{\mathrm{\;o}}}{p},z,\bar{z},\eta^{-})=0\,.
\ee
Thus, taking into account (\ref{sol-2}) for the kernel $\mathcal{A}(\overset{_{\mathrm{\;o}}}{p},z,\bar{z},\ eta^{\pm})$, we have the expression
\be \lb{sol-3a}
\mathcal{A}(\overset{_{\mathrm{\;o}}}{p},z,\bar{z},\eta^{\pm}) \ = \ \delta(\eta^+) \, \delta(z+\bar z) \,
\tilde{\tilde{\mathcal{A}}}(\overset{_{\mathrm{\;o}}}{p},z-\bar{z},\eta^{-})\,,
\ee
Analogously, using the relations (\ref{eq-01}) and (\ref{eq-02}) one gets the only equation:
\be \lb{eq-l}
\left(z-\bar z\right) \frac{\partial}{\partial \eta^{-}} \, \tilde{\tilde{\mathcal{A}}}(\overset{_{\mathrm{\;o}}}{p},z,\bar{z},\eta^{\pm})=
\bm{\rho}\, \tilde{\tilde{\mathcal{A}}}(\overset{_{\mathrm{\;o}}}{p},z,\bar{z},\eta^{\pm})  \, .
\ee
Solution to this equation looks like
\be \lb{sol-3b}
\mathcal{A}(\overset{_{\mathrm{\;o}}}{p},z,\bar{z},\eta^{\pm}) \ = \ \delta(\eta^+) \, \delta(z+\bar z)\,
e^{\,\bm{\rho}\eta^-\!{/}(z-\bar z)} \, f(z-\bar z)\,,
\ee
where $f(z-\bar z)$ is an arbitrary function. After restoring the original variables, the solution (\ref{sol-3b}) is written as (\ref{sol-2b}).

\section*{Appendix\,2: \\
The value of the Casimir operator on the field $\Psi(p,\eta)$ } \lb{sect-a1}
\def\theequation{B.\arabic{equation}}
\setcounter{equation}0

\subsection*{Non-singular case}
Let us find the value of the square of the Pauli-Lubanski vector (\ref{vPL}) on the Lorentz-covariant field $\Psi(p,\eta)$ defined in (\ref{nrez1}).
The generators of the Poincar\'{e} group $\hat{P}_\mu$, $\hat{M}_{\mu \nu}$, acting in the field space $\Psi(p,\eta)$, have the form
\be \lb{genP}
\hat{P}_\mu = p_\mu \, , \;\;\;
\hat{M}_{\mu \nu} = i\Bigl ( p_{\mu} \frac{\partial}{\partial p^{\nu}} - p_{\nu} \frac{\partial}{\partial p^{\mu}} + \eta_{\mu} \frac{\partial}{\partial \eta^{\nu}} - \eta_{\nu} \frac{\partial}{\partial \eta^{\mu}} \Bigr ) \, .
\ee
Substituting these expressions into the square $\hat{W}_\mu \hat{W}^\mu$ of the Pauli-Lubanski vector
(\ref{vPL}), one gets\footnote{In the relation (\ref{kaz}) we already took into account that the fields
$\Psi(p,\eta)$ by construction are eigenvectors of the operators $\hat{P}_n$ with eigenvalues $p_n$. The massless field is also taken into account: $\hat{P}^n\hat{P}_n=0$.}
\be \lb{kaz}
\hat{W}^2 = 2 (p\cdot \eta) \Bigl (p\cdot \frac{\partial}{\partial \eta} \Bigr) \Bigl (\eta \cdot \frac{\partial}{\partial \eta} \Bigr) - ( p \cdot  \eta)^2 \Bigl (\frac{\partial}{\partial \eta} \Bigr )^2 - \eta^2 \Bigl (p \cdot \frac{\partial}{\partial \eta}\Bigr )^2\,.
\ee

To calculate the value of $\hat{W}^2$ on the field $\Psi(p,\eta)$ defined in (\ref{nrez1}), it is
sufficient to calculate the value of $\hat{W}^2$ on the kernel $\mathcal{A}(p,\eta,\varphi)$
given in (\ref{sol3}). To do this, we represent (\ref{sol3}) as follows:
\be \lb{sol5}
\mathcal{A}(p,\varphi,\eta) = e^{i \bm{\mu} \, \mathrm{B}(p,\eta,\varphi)} \, f(\eta \cdot \eta, p \cdot \eta) \, ,
\ee
where we introduced the notation
\be \lb{sol6}
\mathrm{B} (p,\eta,\varphi) =  \eta \cdot \varepsilon_{(1)}(p,\varphi)/(\eta \cdot p)
\ee
One can show that the function $\mathrm{B} (p,\eta,\varphi)$ satisfies the following properties\footnote{In what follows, we omit the arguments of the function $\mathrm{B} (p,\eta,\varphi)$ and just write $\mathrm{B}$. }:
\be \lb{prfP}
\Bigl ( p \cdot \frac{\partial}{\partial \eta} \Bigr) \mathrm{B}  = \Bigl ( \eta \cdot \frac{\partial}{\partial \eta} \Bigr)  \mathrm{B} =  \Bigl ( \frac{\partial}{\partial \eta}  \cdot \frac{\partial}{\partial \eta}  \Bigr)  \mathrm{B} = 0 \,
\ee
and the relation
\be \lb{prfP1}
(p \cdot \eta)^2 \, \Bigl ( \frac{\partial}{\partial \eta} \, \mathrm{B} \Bigr)  \cdot \Bigl ( \frac{\partial}{\partial \eta} \, \mathrm{B} \Bigr)  = - 1\,.
\ee
When proving (\ref{prfP}) and (\ref{prfP1}), we used relations (\ref{sor1}) and (\ref{sor2}).

To simplify further calculations, we introduce the following notation:
\be \lb{obst}
\begin{array}{c}
\displaystyle
f_{\{1\}} := \frac{\partial}{ \partial (\eta \cdot \eta)} \, f(\eta \cdot \eta, p \cdot \eta) \, , \;\;\;
f_{\{2\}} := \frac{\partial}{ \partial (p \cdot \eta)} \, f(\eta \cdot \eta, p \cdot \eta) \, ,  \\[15pt]
\displaystyle
f_{\{1,2\}} := \frac{\partial}{ \partial (p \cdot \eta)} \, \frac{\partial}{ \partial (\eta \cdot \eta)} \, f(\eta \cdot \eta, p \cdot \eta) = \frac{\partial}{ \partial (\eta \cdot \eta)} \, \frac{\partial}{ \partial (p \cdot \eta)} \, f(\eta \cdot \eta, p \cdot \eta) =: f_{\{2,1\}} \, , \\[15pt]
\displaystyle
f_{\{1,1\}} := \frac{\partial}{ \partial (\eta \cdot \eta)} \, \frac{\partial}{ \partial (\eta \cdot \eta)} \, f(\eta \cdot \eta, p \cdot \eta) \, .
\end{array}
\ee
Let us now act on the right-hand side (\ref{sol5}) by the first term of the right-hand side (\ref{kaz}). As a result, we get
\be \lb{tdp1}
 2 (p\cdot \eta) \, e^{i \bm{\mu} \, \mathrm{B}} \, \Bigl (4 \, (\eta \cdot \eta) (p \cdot \eta) \, f_{\{1,1\}}   + 4 (p \cdot \eta) \, f_{\{1\}} + 2 (p\cdot \eta)^2 \, f_{\{1,2\}} \Bigr ) \, ,
\ee
where we have used (\ref{prfP}) and notation (\ref{obst}). Next, let us act by the second term of the
right-hand side (\ref{kaz}) on the right-hand side (\ref{sol5}). As a result, we get
\be \lb{tdp2}
- \bm{\mu}^2 \cdot e^{i \bm{\mu} \, \mathrm{B}} \, f (\eta \cdot \eta, p \cdot \eta) - (p \cdot \eta)^2 \, e^{i \bm{\mu} \, \mathrm{B}} \, \Bigl ( 4 (\eta \cdot \eta) \,  f_{\{1,1\}}  + 4 (p \cdot \eta) \, f_{\{1,2\}} + 8 f_{\{1\}}  \Bigr ) \, ,
\ee
where the identities (\ref{prfP}), (\ref{prfP1}) and notation (\ref{obst}) were again used in the
derivation. Let us act on the right-hand side (\ref{sol5}) by the last term of the right-hand side (\ref{kaz}). The result is written as
\be \lb{tdp3}
-4 \cdot (\eta \cdot \eta) \, e^{i \bm{\mu} \, \mathrm{B}} \, (p \cdot \eta)^2 \, f_{\{1,1\}} \,  \, .
\ee
Summing up (\ref{tdp1}), (\ref{tdp2}), and (\ref{tdp3}), we get the equality
\be \lb{basn1}
\hat{W}^2 \mathcal{A}(p,\eta,\varphi) = - \bm{\mu}^2 \, \mathcal{A}(p,\eta,\varphi) \,
\ee
and hence we have
\be \lb{basn2}
\hat{W}^2 \Psi(p, \eta) = -  \bm{\mu}^2 \,  \Psi(p, \eta) \, .
\ee
The latter shows that the irreducible infinite spin representation of the Poincar\'{e} group
 is realized on the Lorentz-covariant field $\Psi(p, \eta)$.

\subsection*{Singular case}
Let us now calculate the value of $\hat{W}^2$ on the field $\Psi(p,\eta)$ defined by relation
(\ref{nrez1-b}). To do this, it is sufficient to find eigenvalue $\hat{W}^2$ on the kernel $\mathcal{A}(p,\eta,\varphi)$, which in this case is given by  expression (\ref{sol-2d}).

The first two terms in the relation (\ref{kaz}) for $\hat{W}^2$ vanish on the kernel
$\mathcal{A}(p,\eta,\varphi)$ due to the presence of $\delta$-functions in the kernel $\mathcal{A}(p,\eta,\varphi)$ and the property of the $\delta$-function $x \, \delta'(x) = - \delta( x)$. The last term in (\ref{kaz}) acting on $\mathcal{A}(p,\eta,\varphi)$ gives the following result:
\be \lb{skaz-1}
\bm{\mu}^2 \, \frac{\eta \cdot \eta}{\bigl ( \eta\cdot {\varepsilon}_{(1)}(\varphi) \bigr)^2} \, \mathcal{A}(p,\eta,\varphi)\,.
\ee
Since the four vectors $p^m$, $\varepsilon^m$, ${\varepsilon}^m_{(1)}(\varphi)$, ${\varepsilon}^m_{(2)}(\varphi )$
(as well as the four vectors $\overset{_{\mathrm{\;o}}}p^m$, $\overset{_{\mathrm{\;o}}}\varepsilon{}^m$, $\overset{_{\mathrm{\;o}}}{\varepsilon}{}^m_{(1)}(\varphi)$, $\overset{_{\mathrm{\;o}}}{\varepsilon }{}^m_{(2)}(\varphi)$) form an orthonormal basis in the four-dimensional vector space, we have the equality
\begin{eqnarray}
\lb{skaz-2}
\eta \cdot \eta &=& 2 (\eta \cdot p) (\eta \cdot \varepsilon) -  (\eta\cdot {\varepsilon}_{(1)}(\varphi))^2 - (\eta\cdot {\varepsilon}_{(2)}(\varphi))^2 \\
\nonumber
&=& 2 (\eta \cdot \overset{_{\mathrm{\;o}}}{p}) (\eta \cdot \overset{_{\mathrm{\;o}}}{\varepsilon}) -  (\eta\cdot \overset{_{\mathrm{\;o}}}{\varepsilon}_{(1)}(\varphi))^2 - (\eta\cdot \overset{_{\mathrm{\;o}}}{\varepsilon}_{(2)}(\varphi))^2  \, .
\end{eqnarray}
Substituting (\ref{skaz-2}) into (\ref{skaz-1}) and taking into account the presence of $\delta$-functions $\delta(\eta\cdot p)$ and $\delta(\eta\cdot { \varepsilon}_{(2)}(\varphi))$ in expression (\ref{sol-2d}) for the kernel $\mathcal{A}(p,\eta,\varphi)$, we get
\be \lb{skaz-4}
\hat{W}^2  \, \mathcal{A}(p,\eta,\varphi) = - \bm{\mu}^2 \, \mathcal{A}(p,\eta,\varphi) \, .
\ee
The corresponding relativistic field satisfies the same equation:
\be \lb{basn2b}
\hat{W}^2 \Psi(p, \eta) = -  \bm{\mu}^2 \,  \Psi(p, \eta) \, .
\ee
As a result, it describes the irreducible infinite spin representation of the Poincar\'{e} group.

\end{document}